\documentclass[11pt]{article}
\pdfoutput=1
\usepackage{jcapmod}

\usepackage{mathtools}
\usepackage{booktabs}
\usepackage[english]{babel}
\usepackage{amsmath,amssymb,amsbsy,amstext, amsthm, simplewick, amsfonts}
\usepackage{hyperref}
\usepackage{graphicx}
\usepackage[small]{caption}
\usepackage{slashed}
\usepackage{siunitx}
\usepackage{upgreek}
\usepackage{framed}
\usepackage{wrapfig}
\usepackage{multirow}
\usepackage{bbm}
\usepackage[svgnames,dvipsnames,x11names]{xcolor}
\usepackage[utf8x]{inputenc}
\usepackage{selinput}
\usepackage{bm}
\usepackage{float}
\usepackage{geometry}
\usepackage{yfonts}
\usepackage{caption}
\usepackage{subcaption}
\usepackage{sidecap}
\usepackage{longtable}
\usepackage{anyfontsize}
\usepackage{dsfont}
\usepackage{tikz}
\usepackage{relsize}
\usepackage{tcolorbox}

\usepackage{colortbl}

\makeatletter
\newlength{\apb@width}
\newcommand{\autoparbox}[2][c]{\settowidth{\apb@width}{#2}\parbox[#1]{\apb@width}{#2}}

\makeatother

\definecolor{lightgray}{gray}{0.9}

\usepackage[framemethod=default]{mdframed}
\newmdenv[skipabove=7pt,
skipbelow=7pt,
rightline=false,
leftline=false,
topline=false,
bottomline=false,
backgroundcolor=gray!10,
linecolor=gray,
innerleftmargin=5pt,
innerrightmargin=5pt,
innertopmargin=5pt,
innerbottommargin=5pt,
leftmargin=0cm,
rightmargin=0cm,
linewidth=4pt]{eBox}

\numberwithin{equation}{section}

\def\beq{\begin{equation}}
\def\eeq{\end{equation}}

\def\bea{\begin{eqnarray}}
\def\eea{\end{eqnarray}}

\def\d{{\rm d}}

\def\beq{\begin{equation}}
\def\eeq{\end{equation}}
\def\bea{\begin{eqnarray}}
\def\eea{\end{eqnarray}}

\def\d{{\rm d}}

\def\T{{\cal T}}

\def\fnl{f_{\rm NL}}
\def\d{{\rm d}}

\def\k{{\vec k}}

\def\q{{\vec q}}

\def\p{{\vec p}}

\def\x{{\vec x}}

\def\y{{\vec y}}

\def\fnleq{{f_{\rm NL}^{\rm eq}}}
\DeclareRobustCommand{\SkipTocEntry}[4]{}

\def\dgal{{\delta_g}}
\def\dgalO{{\delta^{\rm obs}_g}}
\def\dbarO{{\bar \delta}^{\rm obs}}
\def\knl{{k_{\rm NL}}}

\definecolor{blue3}{RGB}{31, 119, 180}
\definecolor{red3}{RGB}{	214, 39, 40}
\definecolor{orange3}{RGB}{255, 127, 14}
\definecolor{green3}{RGB}{44, 160, 44}

\begin{document}

\begin{titlepage}
\setcounter{page}{1} \baselineskip=15.5pt 
\thispagestyle{empty}

\begin{center}
{\fontsize{24}{18} \bf The Power of Locality:} \\[8pt]
{\fontsize{18}{18} \bf Primordial Non-Gaussianity at the Map Level}
\end{center}

\vskip 20pt
\begin{center}
\noindent
{\fontsize{12}{18}\selectfont  Daniel Baumann$^{1,2,3}$ and Daniel Green$^{4}$}
\end{center}

\begin{center}
\vskip 4pt
\textit{ $^{1}$ Institute of Physics, University of Amsterdam, Amsterdam, 1098 XH, The Netherlands} \\[2pt]
\textit{ $^{2}$ Center for Theoretical Physics, National Taiwan University, Taipei 10617, Taiwan } \\[2pt]
\textit{ $^{3}$ Physics Division, National Center for Theoretical Sciences, Taipei 10617, Taiwan} \\[2pt]
\textit{ $^{4}$ Department of Physics,University of California at San Diego, La Jolla, CA 92093, USA}
\end{center}

\vspace{0.4cm}
 \begin{center}{\bf Abstract}
  \end{center}

\noindent
Primordial non-Gaussianity  is a sensitive probe of the inflationary era, with a number of important theoretical targets living an order of magnitude beyond the reach of current CMB constraints.  Maps of the large-scale structure of the universe, in principle, have the raw statistical power to reach these targets, but the complications of nonlinear evolution are thought to present serious, if not insurmountable, obstacles to reaching these goals.  In this paper, we will argue that the challenge presented by nonlinear structure formation has been overstated.  The information encoded in primordial non-Gaussianity resides in nonlocal correlations of the density field at three or more points separated by cosmological distances.  In contrast, nonlinear evolution only alters the density field locally and cannot create or destroy these long-range correlations.  
This locality property of the late-time non-Gaussianity is obscured in Fourier space and in the standard bispectrum searches for primordial non-Gaussianity. We therefore propose to measure non-Gaussianity in the position space maps of the large-scale structure.  As a proof of concept, we study the case of equilateral non-Gaussianity, for which the degeneracy with late-time nonlinearities is the most severe. 
 We show that a map-level analysis is capable of breaking this degeneracy and thereby significantly improve the constraining power over previous estimates. Our findings suggest that ``simulation-based inference" involving the forward modeling of large-scale structure maps has the potential to dramatically impact the search for primordial non-Gaussianity. 
 
\end{titlepage}

\restoregeometry

\newpage
\setcounter{tocdepth}{2}
\tableofcontents

\newpage

\section{Introduction}

The universe we inhabit is non-Gaussian, as seen in the statistics of cosmological observables like the cosmic microwave background (CMB) and the large-scale structure (LSS).  This non-Gaussianity encodes nonlinear effects ranging from gravitational lensing to the microphysics of structure formation, and measuring it
allows us to reconstruct these important effects in our cosmic history.  Primordial non-Gaussianity in the initial conditions has not yet been observed, but can be a similarly powerful tool in the quest to understand the very early universe~\cite{Chen:2010xka,Meerburg:2019qqi}.  It  probes the interactions~\cite{Alishahiha:2004eh,Chen:2006nt,Moss:2007cv,LopezNacir:2011kk} and the particle content~\cite{Chen:2009zp,Baumann:2011nk,Assassi:2012zq,Noumi:2012vr,Arkani-Hamed:2015bza,Lee:2016vti,Arkani-Hamed:2018kmz} during inflation, which are linked to a variety of inflationary mechanisms. 
Moreover,  since inflation is likely to have occurred at very high energies, primordial non-Gaussianity provides the opportunity of testing physics at energies well beyond the reach of terrestrial experiments~\cite{Arkani-Hamed:2015bza}.  

\vskip 4pt
The rich structure of non-Gaussian correlators is a topic of active exploration, both on purely theoretical grounds and as a target for future observations.  While theoretical progress continues to unearth new questions that could be answered through measurements of primordial non-Gaussianity, improving the measurements themselves remains a fundamental challenge.  So far, the CMB has been the driving force behind these measurements~\cite{Planck:2019kim}, but there are unfortunately not enough modes in the CMB to improve the sensitivity by more than a factor of a few~\cite{Abazajian:2019eic}.  Ultimately, the future of these measurements will therefore rest on LSS surveys~\cite{Alvarez:2014vva}. 

\begin{figure}[t!]
\centering
\includegraphics[scale=1.1]{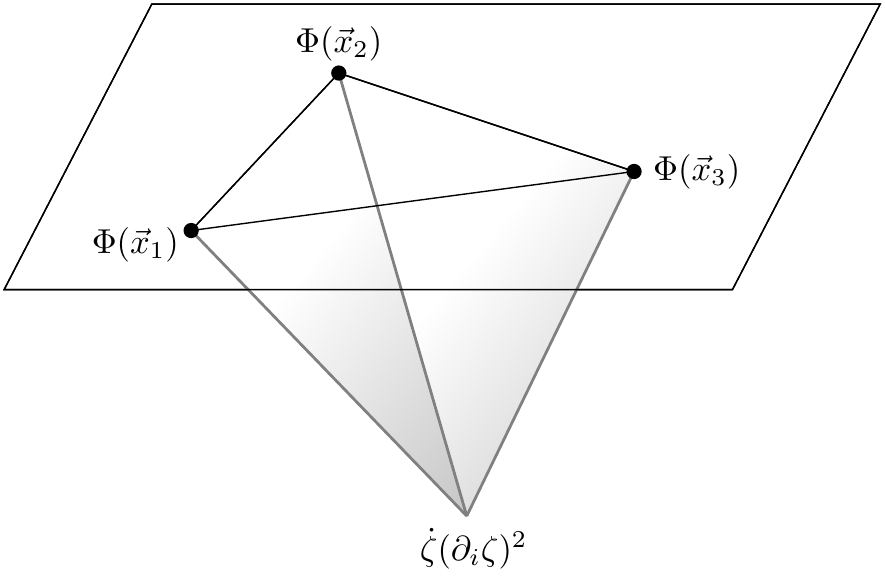}
\caption{Correlations in the quantum fluctuations during inflation get stretched to superhorizon scales, producing  apparently nonlocal correlations after inflation.
By causality, these correlations are protected from the effects of late-time nonlinearities. }
\label{fig:ThreePoint}
\end{figure}

\vskip 4pt
In the case of {\it local non-Gaussanity}~\cite{Komatsu:2001rj}, upcoming surveys like SPHEREx~\cite{Dore:2014cca} have the potential to significantly improve over the CMB constraints 
 because the signal is peaked on large scales~\cite{Dalal:2007cu,dePutter:2014lna}, well away from the nonlinear regime.  
An equally important theoretical target is {\it equilateral non-Gaussianity}~\cite{Babich:2004gb}, which arises from cubic self-interactions during single-field inflation and correlates the initial density fluctuations at three separated points, as illustrated in Figure~\ref{fig:ThreePoint}.  
These interactions are small for slow-roll inflation~\cite{Maldacena2,Creminelli:2003iq}, but can be significant in alternative scenarios~\cite{Alishahiha:2004eh,Chen:2006nt,Moss:2007cv,LopezNacir:2011kk}. In fact, a detection of equilateral non-Gaussianity at the level $\fnleq > 1$ would rule out the slow-roll paradigm and require a different inflationary mechanism~\cite{Baumann:2014cja,Baumann:2015nta}. This makes the measurement of equilateral non-Gaussianity at this level of sensitivity an important long-term goal for future LSS surveys~\cite{Alvarez:2014vva}.

\vskip 4pt
The prospects of a making a competitive measurement of equilateral non-Gaussianity using LSS observations is generally thought to be limited by our ability to model gravitational nonlinearities and galaxy formation in the late universe~\cite{Baldauf:2016sjb,Karagiannis:2018jdt}.  All of these secondary effects contribute to a nontrivial three-point function (or bispectrum) that is highly degenerate with the primordial signal~\cite{Lazanu:2015rta,Lazanu:2015bqo}.  While perturbation theory~\cite{Bernardeau:2001qr} (including the EFT of LSS~\cite{Baumann:2010tm,Carrasco:2012cv,Porto:2013qua}) could be used to model these nonlinearities, one would have to calculate the corrections very accurately to eliminate both biases and degeneracies in the measurement of $\fnleq$.  More realistically, our limitation in calculating these effects to arbitrary precision presents a source of {\it theoretical error} that, in many surveys, would be significantly larger than shot noise or cosmic variance on the scales of interest~\cite{Baldauf:2016sjb}.

\newpage
While the challenge imposed by gravitational nonlinearities may seem daunting, it rests on two implicit assumptions: 
\begin{itemize}
\item First, it assumes that the only way to distinguish primordial physics and late-time astrophysics is accurate and precise modeling of the relevant effects.  However, this does not allow for the possibility of ``protected observables" that are robust to these nonlinearities and can therefore be measured reliably.  Local non-Gaussianity provides a compelling example of this, since it produces a specific scale-dependent bias~\cite{Dalal:2007cu} that cannot be mimicked by late-time dynamics without violating the equivalence principle. In this paper, we will show that part of the signal from equilateral non-Gaussianity is similarly protected. 
\item  Second, it assumes that, because the primordial signal lives in two- and three-point functions, we should project both the signal and the late-time nonlinear contributions onto this limited set of correlation functions.  Unlike the primordial signal, however, the late-time nonlinearities also impact  higher-point correlators whose information is then being neglected.  
This suggests that the primordial and late-time signals may be more distinguishable than one is led to believe from a standard bispectrum analysis. In this paper, we will argue that this information is more manifest in maps of the large-scale structure.
\end{itemize}

\begin{figure}[t!]
\centering
\includegraphics[scale=1.1]{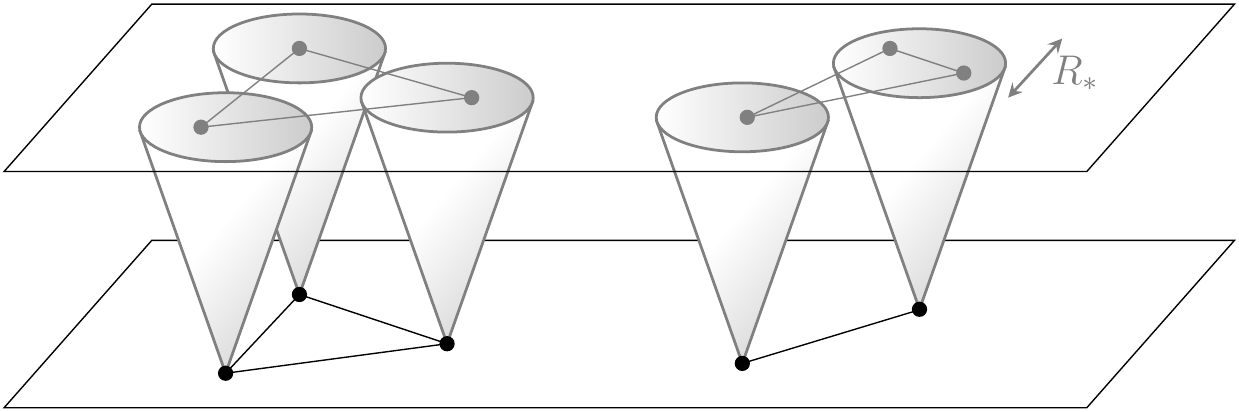}
\caption{Schematic illustration highlighting the difference between primordial non-Gaussianity ({\it left}) and late-time non-Gaussianity ({\it right}).  The non-Gaussianity produced by late-time nonlinearities is constrained by the locality scale $R_*$, while primordial non-Gaussianity gives rise to apparently nonlocal correlations.}
\label{fig:causal}
\end{figure}

\vskip 4pt
 A distinguishing feature of inflation is that it stretches quantum fluctuations to super-Hubble scales, creating superhorizon modes after inflation~\cite{Hu:1996yt,Spergel:1997vq,Dodelson:2003ip}. Local self-interactions during inflation, like those producing equilateral non-Gaussianity, then create apparently nonlocal correlations between distant points, generated at the overlap of their past lightcones~\cite{Green:2020whw} (see Figure~\ref{fig:ThreePoint}).  In contrast, late-time nonlinearities only act locally (see e.g.~\cite{Baumann:2010tm} for a discussion) and cannot alter these nonlocal correlations after inflation is over. 
As illustrated in Figure~\ref{fig:causal}, causality then ensures that the imprints of primordial non-Gaussianity at the map level cannot be created or destroyed by local short-distance physics.\footnote{A similar feature is found in holography where the local bulk physics is nonlocally encoded in the boundary data~\cite{Susskind:1998vk,Polchinski:1999ry,Gary:2009ae,Fitzpatrick:2011ia,Maldacena:2011nz,Raju:2012zr,Maldacena:2015iua}.  In that context, the protection we are describing manifests itself as an error-correcting code~\cite{Almheiri:2014lwa}.}  
 
\vskip 4pt
The local nature of the late-time nonlinearities is obscured when non-Gaussianity is analyzed using only the bispectrum. Instead, the difference between the primordial and late-time signals appears more directly in the properties of the position space map; equivalently, one requires more than just the power spectrum and the bispectrum to see the apparent nonlocality of the inflationary signal in action.  As we will show, it is the information from higher-point correlation functions
that breaks the degeneracies between the primordial non-Gaussianity and the late-time nonlinearities (see~\cite{Cole:2020gzt} for another approach to finding protected non-Gaussian properties of these maps). Concretely, at cubic order, the nonlinearities affect not just the bispectrum, but also the trispectrum, which is being missed as an additional constraint in a bispectrum-only analysis. These relations between different correlation functions are required by locality and are hardwired in the map-based analysis.

\vskip 4pt
The strategy we will advocate in this paper
 is closely aligned with ongoing work in LSS theory and analysis.  Several groups have been using field-level likelihoods of the dark matter, halos and galaxies 
 in order to test our modeling beyond the predictions of individual correlation functions~\cite{Schmittfull:2018yuk,Cabass:2019lqx,Cabass:2020nwf,Schmidt:2020viy,Schmittfull:2020trd}. 
 In parallel, there is a large effort to apply ``simulation-based inference" (made feasible by the application of machine learning) to cosmological data analysis~\cite{Schmittfull:2017uhh,Taylor:2019mgj,Dai:2020ekz,Modi:2020dyb,Modi:2021acq,Makinen:2021nly,Hassan:2021ymv,Villaescusa-Navarro:2021pkb,Villaescusa-Navarro:2021cni}. 
 By  forward modeling the cosmological maps, these approaches try to sample realizations of the initial conditions (and the cosmological parameters) and find those that best reproduce the observed maps of the late universe.  In principle, these methods can produce the map-level constraints on primordial non-Gaussianity that we will describe analytically.  However, the simulations are inevitably incomplete and one might worry that the nonlinear physics which is not included (or not simulated at sufficient accuracy) fundamentally limits the effectiveness of these approaches, given the very small size of the signals.  Our main goal is to demonstrate that the map-based data analysis is much more sensitive to primordial non-Gaussianity than the standard bispectrum analysis and that this does not require perfect forward modeling as long as the physics involved is local in space.  

\vskip 4pt
The plan of the paper is as follows: In Section~\ref{sec:intuitive}, we explain in more detail why late-time nonlinearities are the main challenge for extracting primordial non-Gaussianity from LSS observations and argue that the apparent nonlocality of the inflationary signals can be used to extract them reliably from position space maps.  In Section~\ref{sec:map}, we quantify the degeneracies between the primordial and late-time signals by deriving the Fisher matrices associated to a map-level analysis with and without cosmic variance.  We show that only quadratic nonlinearities affect the map-level analysis, while all higher-order nonlinearities decouple.  In Section~\ref{sec:forecasts}, we forecast the expected sensitivities of the map-level analysis and compare them to the results of the standard bispectrum analysis. Our conclusions are presented in Section~\ref{sec:conclusions}.  Appendix~\ref{app:likelihoods} contains the derivations of the maximum likelihood maps and the resulting Fisher matrices.

\paragraph{Notation and conventions} \ \\ 
Most of our notation will be introduced as we go along. Here, we just point out that we
will invoke several different density contrasts, as listed in the following table:
\begin{center}
\begin{tabular}{ll}
\toprule
$\delta$ & linearly evolved matter density contrast \\
$\bar \delta$ & specific realization of $\delta$\\
$\bar \delta^{\rm obs}$ & maximum likelihood solution of $\delta$ \\
$\delta_{\rm G}$ & Gaussian part of $\delta$ \\
$\delta_{\rm NG}$ & non-Gaussian part of $\delta$ \\
$\delta_g$ & galaxy density constrast (model) \\
$\delta_g^{\rm obs}$ & galaxy density constrast (observed)  \\
\bottomrule
\end{tabular}
\end{center}
To avoid clutter, we often drop the subscript on $\delta_{\rm G}$, in which case the context must be used to distinguish it from the full density contrast $\delta$. 
We will use $\delta_{\rm D}(\x - \x^{\hskip 1pt \prime})$, with the subscript, to denote the Dirac delta function.

\newpage
\section{Non-Gaussianity with LSS}
\label{sec:intuitive}

A detection of primordial non-Gaussianity would be a remarkable opportunity to learn about the physics of inflation. 
Unfortunately, the expected signals are small and hard to extract from cosmological data. In this section, we first  review the promise of non-Gaussianity as a probe of inflation and then explain why measuring the inflationary signals with LSS observations is so challenging. 
We describe the pessimistic view that gravitational nonlinearities in the late universe are hard to characterize and produce bispectra that are hard to distinguish from popular forms of inflationary non-Gaussianity.
We then show that the apparent nonlocality of the inflationary signals is a protected feature that cannot be mimicked by local effects in the late universe. We are therefore led to the more optimistic view that a position space analysis of LSS maps is a robust way discriminate primordial non-Gaussianity from the secondary non-Gaussianity created at late times.

\subsection{The Promise}

For adiabatic fluctuations, the initial conditions are described by a single degree of freedom, which is typically taken to be the comoving curvature perturbation, $\zeta$, or the Newtonian potential, $\Phi$. During the matter era, we have $\Phi = -\frac{3}{5} \zeta$ on large scale.  
Given the near Gaussianity of the initial conditions inferred from the CMB observations, the leading non-Gaussian statistic is the three-point function of the Fourier modes (or the bispectrum):
\beq
\langle \Phi(\k_1) \Phi(\k_2) \Phi(\k_3) \rangle = B_\Phi(k_1,k_2, k_3) \, (2\pi)^3 \delta_{\rm D}(\k_1+\k_2+\k_3) \, ,
\eeq
with the momentum dependence of the function $B_\Phi(k_1, k_2, k_3)$ encoding much of the physics of the inflationary era, such as the spectrum of particles and their interactions~\cite{Arkani-Hamed:2018kmz}.  

\vskip 4pt
One of the earliest phenomenological models of non-Gaussianity was the so-called
{\it local non-Gaussianity}~\cite{Komatsu:2001rj}, which is generated by writing the Newtonian potential as 
\beq
\Phi(\x) = \varphi(\x) +\fnl^{\rm loc} \left(\varphi^2(\x) -\langle \varphi^2(\x)\rangle \right) ,
\eeq
where $\varphi$ is a Gaussian random field. 
The corresponding bispectrum is
\beq
 B_\Phi^{\rm loc}(k_1,k_2,k_3) = 2 \fnl^{\rm loc} P_\varphi(k_1) P_\varphi (k_2) + {\rm perms} \, , 
\eeq
where $P_\varphi (k)$ is the power spectrum of $\varphi(\k)$.  The current Planck constraint on local non-Gaussianity is $\fnl^{\rm loc} = 2.7 \pm 5.4$ (68\%\,CL)~\cite{Planck:2019kim}.
For a nearly scale-invariant power spectrum, $P_\varphi (k) \propto k^{-3}$, this bispectrum signal peaks in the squeezed limit, $k_3 \ll k_1 \approx k_2$~\cite{Babich:2004gb}.
Such local non-Gaussianity can only arise in multi-field models of inflation and is therefore an important diagnostic of extra degrees of freedom during inflation~\cite{Creminelli:2004yq}.  We can understand this intuitively by noticing that if there is only a single degree of freedom, then $\varphi$ is the Newtonian potential in the Gaussian limit.  A mechanism that generates local non-Gaussianity from the initial Gaussian field is therefore sensitive directly to the Newtonian potential, which is inconsistent with the equivalence principle.  This apparent violation of the equivalence principle survives at late times as a ``scale-dependent bias"~\cite{Dalal:2007cu}, which arises because galaxies form at peaks of the potential $\Phi$ and not the density $\delta \rho \propto \nabla^2 \Phi$.  For the same reason, this scale-dependent bias cannot be mimicked by nonlinear gravitational evolution and is expected to lead to significant improvements in the measurement of $\fnl^{\rm loc}$ in the next decade~\cite{Dore:2014cca}.

\vskip 4pt
In this paper, we revisit the prospects of using LSS observations to look for {\it equilateral non-Gaussianity}, for which the bispectrum signal is dominated by equilateral configurations~\cite{Babich:2004gb}, with $k_1 \approx k_2 \approx k_3$.  This is the dominant non-Gaussianity in models of single-field inflation, with cubic interactions of the form $\dot \zeta^3$ and $\dot \zeta (\partial_i \zeta)^2$.  For concreteness, we will focus on the equilateral shape coming from the $\dot \zeta^3$ interaction,
 \beq
B_\Phi^{\rm eq}(k_1,k_2,k_3) = 162 \, \fnleq \,\frac{ \Delta_\Phi^2}{k_1 k_2 k_3 (k_1+k_2 +k_3)^3} \, ,
\label{equ:Beq}
\eeq
where $\Delta^2_\Phi \equiv (k^3/2\pi^2) P_\Phi(k)$, but other forms of equilateral non-Gaussianity are very similar.  An important feature of this bispectrum shape is the pole at vanishing total energy $k_t \equiv k_1 +k_2 +k_3 \to 0$.  This pole is a universal feature of local interactions in the early-time limit of inflation~\cite{Raju:2012zr, Maldacena:2011nz, Arkani-Hamed:2018kmz}.  It arises because energy is not conserved in a cosmological background, so instead of an energy conserving delta function, the correlator inherits this pole. The order of the pole depends on the number of derivatives of the bulk interaction.  Finally, the appearance of the total energy pole is the Fourier space manifestation of mode coupling at the past intersection of the lightcones of three separated points~\cite{Green:2020whw}.

\vskip 4pt
Observations of the CMB anisotropies have put interesting constraints on the amplitude of equilateral non-Gaussianity, $\fnleq =  -26 \pm 47$ (68\%\,CL)~\cite{Planck:2019kim}.  While these observations have reached an impressive level of precision, the inferred limits on equilateral non-Gaussianity are still more than an order of magnitude from an interesting threshold value of $\fnleq = O(1)$~\cite{Baumann:2014cja}. If the inflationary fluctuations are weakly coupled excitations of a strongly coupled background --- like the pions in QCD --- then they naturally produces equilateral non-Gaussianity with $\fnleq \gtrsim 1$. If, on the other hand, inflation was of the slow-roll type with only perturbative higher-derivatives interactions of the inflaton field, then we expect $\fnleq \lesssim 1$.
Getting to $\fnleq = O(1)$ is therefore a natural target for future observations. However, reaching this level of sensitivity will be an enormous challenge.

\subsection{The Challenge} \label{sec:challenge}

Unfortunately, the universe does not let us observe the primordial perturbations ($\zeta$ or $\Phi$) directly, but instead their statistical properties must be inferred from the statistics of inhomogeneities and anisotropies of late-time cosmological observables.  While the anisotropies of the CMB have provided an incredibly powerful window into inflation, reaching our aspirational goals will require measuring non-Gaussanity through tracers of the density fluctuations in the late universe.  

\vskip 4pt
The linearly evolved density field is related to the initial conditions by 
\beq
\delta(\k,z) = \frac{2 k^2 T(k) D(z)}{3  \Omega_m H_0^2}\, \Phi(\k)  \equiv \T(k,z) \hskip 1pt \Phi(\k)  \, ,
\label{equ:deltaDEF}
\eeq
where $T(k)$ is the linear transfer function and $D(z)$ is the linear growth function at redshift $z$. 
In the following, all correlations will be evaluated at a fixed redshift and we will drop the explicit redshift dependence to avoid clutter.  The linear density field inherits the correlations of the primordial fluctuations, 
\begin{align}
P(k)  &\equiv \langle \delta(\k) \delta(\k') \rangle' = \T^{\hskip 1pt 2}(k) P_\Phi(k)\,, \\
B(k_1,k_2,k_3) &\equiv   \langle \delta(\k_1) \delta(\k_2) \delta(\k_3) \rangle' = \T(k_1) \T(k_2) \T(k_3)  B_\Phi(k_1,k_2,k_3) \, ,
\end{align}
where the prime on the expectation values denotes dropping the momentum-conserving delta function, $(2\pi)^3 \, \delta_{\rm D}(\sum \k_i)$. 
At times, it will be useful to approximate the linearly evolved matter power spectrum by  
\beq\label{eq:power_law}
P(k) =  \frac{2 \pi^2 }{k_{\rm  NL}^3 } \left(\frac{k}{\knl} \right)^{\Delta}\,,
\eeq
where $\knl \approx 0.3 \, h \, {\rm Mpc}^{-1}$ and $\Delta \in (-1, -2)$ is a good fit   at $z=0$   and in the range $k \in (0.1,0.25) \, h \, {\rm Mpc}^{-1}$.  At larger redshift,  modes with $k> 0.25 \, h \, {\rm Mpc}^{-1}$ may be accessible and $\Delta < -2$ may be appropriate.

\subsubsection*{Biasing}

Most of the information in the late universe is {\it not} encoded in linear modes, but is significantly impacted by nonlinear evolution.  One manifestation of this is the nonlinear galaxy (or halo) biasing that enters late-time observables.  Because these collapsed objects form locally in peaks of the underlying matter density, the galaxy density contrast can be written as a double expansion in fluctuations
and spatial derivatives
\beq
\begin{aligned}
\dgal(\x) &= \sum b_{\cal O} {\cal O} \\[4pt]
&= b_1 \delta + b_2 \delta^2 +  b_3 \delta^3 + \cdots  + b_{\partial^2} R_*^2 \partial^2 \delta + \cdots \, ,
\end{aligned}
\label{equ:biasing}
\eeq
where $\delta$ is the {\it linear} matter density contrast.
In general, this bias expansion should include all terms allowed by the symmetries, which can be constructed from the tidal tensor $\partial_i \partial_j \Phi$~\cite{McDonald:2009dh, Assassi:2014fva}. At quadratic order, this includes the operator ${\cal G}_2 \equiv (\partial_i \partial_j \Phi)^2 - (\partial^2 \Phi)^2$ (with associated bias parameter $b_{{\cal G}_2}$), while at third order, there are three more operators (at leading order in derivatives) that are not written explicitly in (\ref{equ:biasing}). Having said that, the operators $\delta^n$ are sufficient to understand the challenges associated with late-time nonlinearity.  For simplicity of presentation, we will therefore often drop the additional terms and study the simplified biasing model shown in \eqref{equ:biasing}.  We will re-introduce the additional operators in Section~\ref{sec:forecasts} when we present detailed forecasts. 

\vskip 4pt
We have shown one representative higher-derivative operator $R_*^2 \partial^2 \delta$. The length scale $R_*$ appearing in this operator depends on the scale over which the matter evolves, thus setting the scale on which halo and galaxy formation is nonlocal. For dark matter fluctuations, the locality scale $R_*$ is of order the nonlinear scale $k_{\rm NL}^{-1}$ where perturbation theory breaks down.   A well-defined biasing expansion therefore only applies to scales larger than $R_*$. In fact, in order for the double expansion in (\ref{equ:biasing}) to be well-defined, it should be written in terms of {\it renormalized operators}~\cite{McDonald:2009dh, Assassi:2014fva}
\beq
\dgal(\x) = \sum b_{\cal O}^{(R)} [{\cal O}] \, .
\label{equ:biasing2}
\eeq
Defining renormalized operator can be difficult in general, but when $\delta$ is a Gaussian random field, the renormalization procedure simply means subtracting all self-contractions from the composite operators~\cite{Assassi:2014fva}. For example, for the quadratic operator $\delta^2$, we subtract the variance, $[\delta^2] = \delta^2 - \langle \delta^2 \rangle$. The Fourier transform of the renormalized operators, $[\delta^n](\k)$, is simply a convolution of $n$ Fourier modes $\delta(\k)$.  
Beside making the expansion in (\ref{equ:biasing}) well-defined, renormalized operators define an orthogonal basis of operators, e.g.~$\langle [\delta^n](\k) [\delta^m](\k') \rangle \propto \delta_{n,m}$. Working with renormalized operators  therefore removes most off-diagonal correlations, which will play an important role in our discussion. In the following, we will always work with renormalized operators and drop the superscript $(R)$ on the renormalized biasing parameters, $b_{\cal O}^{(R)} \to b_{\cal O}$.

\vskip 4pt
Another point worth highlighting is that the description in (\ref{equ:biasing}) doesn't just capture galaxy biasing, but also encodes the nonlinear evolution of the dark matter itself. It was shown in~\cite{Schmittfull:2018yuk} that, up to bulk flows described by the Zel'dovich approximation, 
the time evolution simply shifts some of the biasing coefficients $b_{\cal O}$ by a calculable amount.  The biasing expansion in~(\ref{equ:biasing}) is therefore a complete representation of the challenge imposed by short-distance gravitational nonlinearities.\footnote{We will neglect the scale-dependent bias~\cite{Dalal:2007cu} associated with $\fnleq$, which appears as a nonlocal effect in this expansion~\cite{LoVerde:2007ri,Matarrese:2008nc,McDonald:2008sc,Desjacques:2010jw,Schmidt:2010gw,Shandera:2010ei,Giannantonio:2011ya,Scoccimarro:2011pz,Baumann:2012bc,Norena:2012yi,Agarwal:2013qta,Assassi:2015fma,Assassi:2015jqa}, because it is doesn't produce competitive constraints on $\fnleq$~\cite{Gleyzes:2016tdh}. }

\vskip 4pt
Even without primordial non-Gaussianity, the nonlinear biasing which defines $\dgal$ will give rise to non-Gaussian correlations.  For example, the galaxy bispectrum is 
\begin{align}
\langle  \dgal(\k_1) \dgal(\k_2) \dgal(\k_3) \rangle  =&\ b_2 b_1^2\left( \langle [\delta^2](\k_1) \delta(\k_2) \delta(\k_3)\rangle + {\rm perms} \right)+ b_2^3\,  \langle [\delta^2](\k_1) [\delta^2](\k_2) [\delta^2](\k_3)\rangle \nonumber \\
&+ b_3 b_2 b_1\left( \langle [\delta^3](\k_1) [\delta^2](\k_2) \delta(\k_3)\rangle+ {\rm perms} \right) + \cdots\,,
\label{equ:XX}
\end{align}
where the ellipses denote higher orders in $\delta$.  In general, all biasing coefficients will be nonzero, $b_n \neq 0$, and  contribute to the bispectrum.
Herein lies the challenge: the coefficients $b_n$ are essentially unknown, as they are determined by the  complex small-scale physics that underlies the formation of galaxies.  While some aspect of these nonlinear terms can be understood theoretically, in general, we cannot even simulate all of the effects that would be needed to determine these parameters from first principles.  As such, the same data we wish to use to measure primordial non-Gaussianity must also be used to determine these bias coefficients (and nonlinear effects more generally).

\subsubsection*{Degeneracies}

At this level, it would appear that our only hope is that the bispectra due to primordial non-Gaussianity are sufficiently distinct from those created by late-time nonlinearities to allow both to be measured simultaneously.  A useful measure of the degeneracy between two bispectra is the inner product~\cite{Babich:2004gb}
\beq\label{eq:bispectrum_dot}
B_i \cdot B_j \equiv V \int \frac{\d^3 k_1 \d^3 k_2 \d^3 k_3}{(2\pi)^9} \frac{B_i(k_1,k_2,k_3) B_j (k_1,k_2,k_3)}{P(k_1) P(k_2) P(k_3)} (2\pi)^3 \delta_{\rm D}(\k_1+\k_2+\k_3)\, ,
\eeq
which is weighted by the signal-to-noise of the bispectrum measurement, assuming that cosmic variance is the dominant source of noise.  We can then also define the so-called ``cosine" of the bispectrum overlap as
\beq\label{eq:cosine_B}
\cos (B_i, B_j) \equiv \frac{B_i \cdot B_j}{\sqrt{B_i \cdot B_i \, B_j \cdot B_j}}\,, 
\eeq
which quantifies how easy it is to distinguish two bispectra $B_i$ and $B_j$ (or how degenerate the amplitudes of the bispectra will be if we measure them simultaneously).  Only if $\cos (B_i, B_j) \ll 1$ are the bispectra $B_i$ and $B_j$ easy to distinguish in a bispectrum analysis.   

\vskip 4pt
Unfortunately, the bispectra from single-field inflation and from nonlinear evolution are both fairly smooth functions and thus generally have large overlaps.  
Moreover, the degeneracy worsens as more unknown biasing parameters are added, meaning that the bispectrum analysis is likely to produce increasingly large errors for $\fnleq$ as more $b_n$ parameters are included.  This has been verified using the modal decomposition of the bispectrum of nonlinear fluctuations in N-body simulations~\cite{Lazanu:2015rta, Lazanu:2015bqo} and is expected to persist for the additional contributions that we cannot yet simulate well, such as galaxy formation and baryonic effects.

\vskip 4pt
Quantitatively, the cosines between the bispectra generated by $b_2$, $b_3$ and $b_4$ and the equilateral bispectrum $B_{\rm eq}$ are  
\begin{align}
\cos(B_{2}, B_{\rm eq}) & \approx  0.96\,, \\
\cos(B_{3}, B_{\rm eq}) & \approx  0.87 \,,\\
\cos(B_{4}, B_{\rm eq}) & \approx 0.80  \, .
\end{align}
These cosine tell us how much $\sigma(\fnleq)$ is impacted by the degeneracy with $b_n$ assuming all other bias parameters are known (i.e.~we truncate the Fisher matrix to two parameters, $b_n$ and $\fnl$).  When $\cos(B_{n},B_{\rm eq})$ is close to unity, the forecasted error scales as 
\beq
\sigma(\fnleq) \approx 
\frac{1}{\sqrt{1- \cos(B_{n}, B_{\rm eq})}}\, \sigma(\fnleq)_{\rm min} \, ,
\eeq
where $\sigma(\fnleq)_{\rm min} =  1/\sqrt{F_{\rm eq, eq}}$ is the optimal measurement of $\fnleq$ set by the {\it Cramer--Rao bound}.  As a result, the degeneracy with $b_2$ alone is expected to increase the error to $\sigma(\fnleq) \approx 5\hskip 1pt \sigma(\fnleq)_{\rm min}$.  Naturally, one would then worry that the cumulative effect of the multitude of nonlinearities that impact structure formation (including those that aren't modeled) will degrade the measurement of $\fnleq$ much further.

\vskip 4pt
In addition to weakening the constraints on $\fnleq$, these degeneracies also suggest that nonlinearity can significantly bias our results.  Inaccurate modeling of nonlinear effects might cause us to over or underestimate the degeneracy with $\fnleq$ and thus could lead to both false positives and negatives.  
Since the nonlinear physics of baryons and galaxy formation is difficult to simulate accurately, this is also a serious concern.

\subsubsection*{Theoretical errors}

Naturally, one might hope to simply avoid these nonlinear corrections by working on large scales where the nonlinearities are small. In that case, perturbation theory allows the precise form of the nonlinear corrections to be calculated order by order in the density contrast~\cite{Bernardeau:2001qr}. Moreover, the  composite operators in the bias expansion lead to momentum integrals in Fourier space, so that the contributions to the correlation functions can be organized in a ``loop expansion".  Any unknown short-distance physics is encoded in the bias parameters (including the effects of nonlinear evolution~\cite{Schmittfull:2018yuk}) and
can be described systematically by the EFT of LSS~\cite{Baumann:2010tm,Carrasco:2012cv,Porto:2013qua}.  This strategy has been successfully employed on current BOSS data to measure cosmological parameters using the power spectrum~\cite{DAmico:2019fhj,Troster:2019ean,Philcox:2021kcw} and bispectrum~\cite{Gil-Marin:2014sta,Slepian:2015hca,Gil-Marin:2016wya,Gil-Marin:2014baa,Philcox:2021kcw}.  In principle, one can imagine measuring successively higher-order correlators~\cite{Philcox:2021eeh,Philcox:2021hbm} to extract additional information from the maps and we anticipate that higher-point information will become  increasingly valuable with larger surveys~\cite{Hahn:2019zob,Oddo:2021iwq,Ivanov:2021kcd}. 

\vskip 4pt
In practice, this approach usually isolates a small number of correlation functions---like the power spectrum and the bispectrum---and calculates them to some fixed order in the loop expansion.  On sufficiently large scales, the errors in such a calculation can easily reach sub-percent levels~\cite{Scoccimarro:2000sn,Carrasco:2013mua,Angulo:2014tfa,Konstandin:2019bay,Steele:2021lnz,Baldauf:2021zlt}.  However, since the number of independent modes decreases on large scales, the regime where perturbation theory is under the best control is also where the data has limited constraining power. Moreover, since the signal becomes weaker on these large scales, it also increases our demands on the accuracy of the model for the nonlinear corrections and any modeling errors become a more serious concern.  The problem can be quantified by treating the uncertainty in the parameters $b_n$ as a source of ``theoretical error"~\cite{Baldauf:2016sjb} (see also \cite{Welling:2016dng}).
For example, the noise covariance matrix for the power spectrum in a fixed redshift survey of volume $V$ is
\beq
C^{-1}_{k k^{\prime}}=\frac{(2 \pi)^{3}}{V} \frac{1}{2 \pi k^{2} \d k}\left(P_{g}(k)+\frac{1}{\bar n}\right)^{2} \delta_{k k^{\prime}}+ (C_{e})_{k k^{\prime}}\,,
\eeq
where $\bar n$ is the number density of objects, which sets the shot noise.  The theoretical error is defined by $\left(C_{e}\right)_{k k^{\prime}}$ which estimates the error in the power spectrum that is being made by truncating the bias expansion~(\ref{equ:biasing2}) at a certain order~\cite{Baldauf:2016sjb}.  
We use the designation ``$b_2$ error" and ``$b_3$ error" to refer to the errors made in assuming linear and quadratic biasing, respectively.\footnote{The $b_2$ and $b_3$ errors are equivalent to the one- and two-loop errors in~\cite{Baldauf:2016sjb}.  The connection to ``loops" can be understood from the one and two additional integrations over momenta needed to calculate the $b_2^2$ and $b_3^2$ contributions to the power spectrum.}
We will assume that these error are uncorrelated in $k$, so that $(C_e)_{k k^{\prime}} \equiv C^2_e(k)\,  \delta_{kk'}$, with 
\beq \label{eq:C_e}
C_e(k) \approx  \left\{\begin{array}{ll}
 b_2^2  \langle [\delta^2](\k) [\delta^2](-\k)  \rangle'  \approx  P(k)  (\hat k/0.31)^{1.8}   &\qquad \text{$b_2$ error} \,,\\[6pt]
 b_3^2  \langle [\delta^3](\k) [\delta^3](-\k)  \rangle'  \approx P(k)  (\hat k/0.23)^{3.3}   & \qquad \text{$b_3$ error}\,,
\end{array}\right.
\eeq
where $\hat k \equiv k/h\hskip 1pt {\rm Mpc}^{-1}$. In the second equality, we have approximated $\int \d^3 p \, P(p) P(|\k-\p\hskip 1pt |) \propto (k/\knl)^{3+\Delta} P(k)$ and set $b_2 \approx b_3 \to 1$ for simplicity.
We have estimated the errors using the power law ansatz in~(\ref{eq:power_law}) with choices of $\Delta$ and $\knl$ chosen to match the errors defined in~\cite{Baldauf:2016sjb}, while still evaluating $P(k)$ as the $z=0$ matter power spectrum in $\Lambda$CDM.  

\vskip 4pt
One key motivation for introducing theoretical error is that it reduces our concerns about a biased measurement of $\fnleq$~\cite{Baldauf:2016sjb}.  If we know the approximate amplitude of the effects that cannot be modeled accurately, we can include them in the noise and thus we down-weight the influence of data where this additional source of noise is large.  This will necessary inflate our error bars, compared to a conventional forecast, since we are neglecting information from modes that previously had high signal-to-noise. In essence, the cost of minimizing biases is an increase in the errors bars.

\vskip 4pt
An important feature of the theoretical error, $\left(C_{e}\right)_{k k^{\prime}}$, is that it is {\it independent} of the volume of the survey.  As a result, even if we increase the number of linear modes by increasing $V$, we will not gain information in proportion to $V$ because it shifts the scale where the theoretical error dominates.  To quantify this, we define the ``effective volume" of a survey as the number of high signal-to-noise modes:
\beq\label{eq:Veff_def}
V_{\rm eff}  \equiv \frac{3}{k_{\rm max}^3} \sum_{k} \frac{P^2(k)}{(V k^2 \Delta k)^{-1} \left(P(k)+ 1/\bar n \right)^2 + C_e(k) } \ ,
\eeq
where $k = n \Delta k$ and $\Delta k = 2\pi / V^{1/3}$.  Taking $C_{e} \to 0$ and $\bar n \to \infty$, we get $V_{\rm eff} \to V$ as expected.  On the other hand, setting $C_e = 0$, but keeping $\bar n$ finite, this definition reduces to the usual definition of the effective volume in the presence of shot noise.  

\begin{figure}[t!]
\centering
\includegraphics[scale=1.]{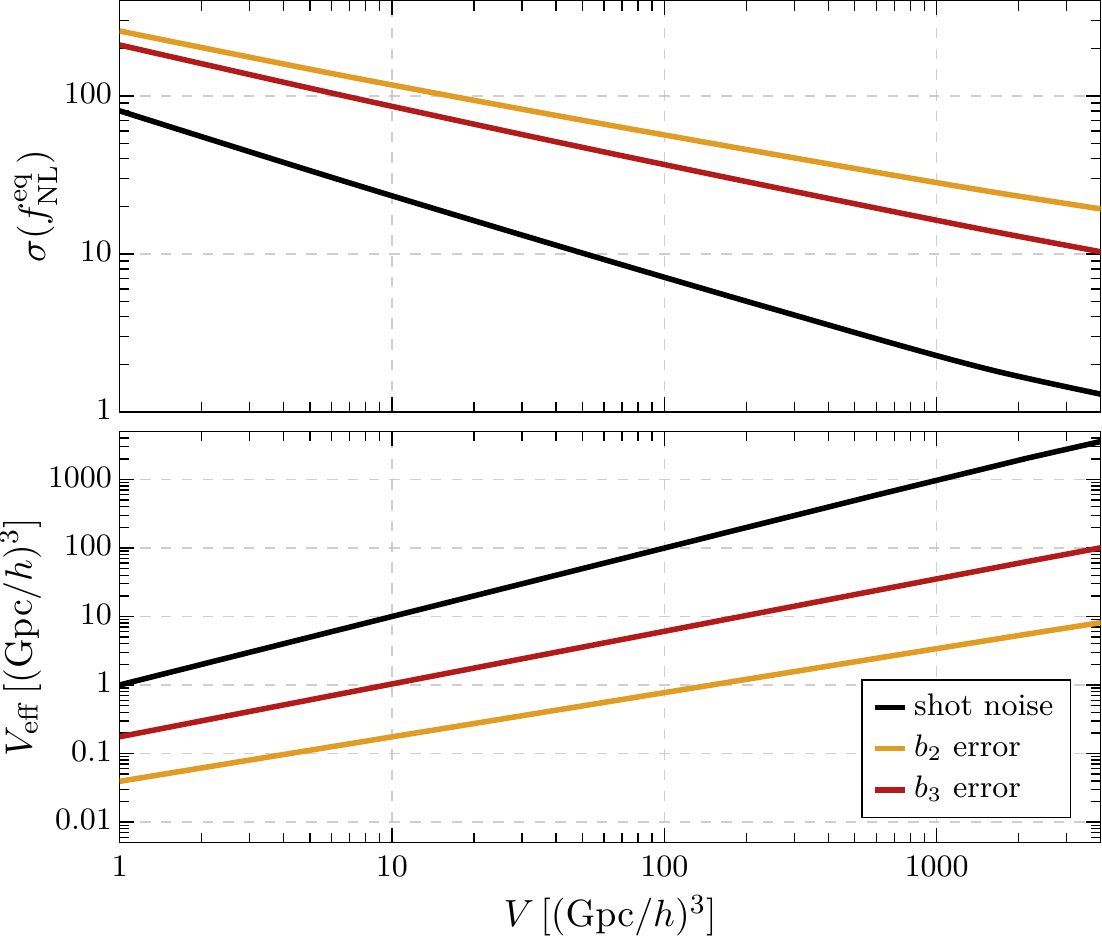}
\caption{Plots of the effective volume ({\it bottom}) and forecasts for $\sigma(\fnleq)$ ({\it top}) as a function of the survey volume, for a $z=0$ survey with $10^{10}$\,objects and $k_{\rm max} = 0.2 \, h \, {\rm Mpc}^{-1}$. Shown are the case of only shot noise ({\it black}), and with $b_2$ ({\it orange}) and $b_3$ ({\it red}) theoretical errors added.  We note that the effective volume does not scale linearly with $V$ in the presence of the theoretical errors. Relatedly, the constraints on $f_{\rm NL}^{\rm eq}$ decrease less rapidly than $1/\sqrt{V}$ when the theoretical errors are included.  We caution the reader not to over-interpret the size of $\sigma(\fnleq)$ in this figure. A realistic survey would include higher redshifts where the universe is more linear and the constraints on $\fnleq$ improve.}
\label{fig:Veff}
\end{figure}

\vskip 4pt
The effective volume at $z=0$ in the presence of theoretical errors is shown in Figure~\ref{fig:Veff}.  We see that the $b_2$ and $b_3$ errors drastically reduce the effective survey volume, suggesting that it will be extremely difficult (if not impossible) to make large improvements in the measurement of $\fnleq$ when the uncertainties in $b_2$ and $b_3$ are accounted for.  As we anticipated, in the presence of theoretical error, $V_{\rm eff}$ does not scale linearly with $V$, leading to the reduced slope of the associated curves in Figure~\ref{fig:Veff} when compared to the result with shot noise alone. In addition, the forecasts for $\fnleq$ scale like $V_{\rm eff}^{-1/2}$ (using the methods from Section~\ref{sec:forecasts} and holding all other parameters fixed) and therefore improve more slowly than $V^{-1/2}$ in the presence of theoretical error.

\vskip 4pt
We are working at $z=0$ to show the full impact of nonlinearity in our universe.  Of course, a realistic survey cannot reach these volumes without going to higher redshifts (if at all) where the universe is more linear. This means that the outlook for realistic surveys is more optimistic than indicated in the figure.  However, in  realistic surveys the exact scale of the nonlinearities can depend sensitively on the observable and can be underestimated by simply rescaling $\knl$. We therefore prefer to attack the problem of nonlinearity head-on at $z=0$.  
\vskip 4pt
One is therefore led to the conclusion that accurate modeling of all nonlinear effects at very high precision is a necessary requirement for significantly improving the measurement of $\fnleq$ with large-scale structure.  The nonlinear bispectra are highly degenerate with our signal; if not modeled correctly they can bias our results.  If we account for the limits of our modeling with theoretical errors, the effective noise levels of future surveys is large which seems to make improving constraints on primordial non-Gaussianity extremely challenging.

\subsection{The Power of Locality}

While the above argument is concerning, it implicitly presumes that the only path to distinguishing the primordial bispectrum from that created by nonlinear evolution is high-precision modeling of the nonlinear universe.  This is particularly worrisome if one considers the wealth of physical effects that we typically do not (or cannot) model accurately (like feedback from baryons). In this paper, we will show that as long as the physics is effectively {\it local}, we can tolerate much more uncertainty than described above, provided we include the information available in the  {\it position space maps} and not just the bispectra measured from those maps.  

\vskip 4pt
To understand the power of locality, let us consider the probability distribution ${\cal P}(\Phi)$ for the primordial potential $\Phi(\x)$ generated by inflation.  This can be written as
\beq
\begin{aligned}
{\cal P}(\Phi)= &\ \exp\bigg[ - \int \d^3 x_1 \d^3 x_2\, \Phi(\x_1) \,C_{2,\Phi}(\x_1-\x_2) \,\Phi(\x_2)\\
&\ + \fnl \int \d^3 x_1 \d^3 x_3 \d^3 x_3\, \Phi(\x_1) \Phi(\x_2) \Phi(\x_3) \,C_{3,\Phi}(\x_1-\x_3,\x_2-\x_3) +\cdots \bigg] \, ,
\end{aligned}
\label{equ:219}
\eeq
where
\begin{align}
C_{2,\Phi}(\x) &= \int \frac{\d^3 k}{(2\pi)^3}  \frac{e^{-i\k\cdot \x}}{P_\Phi(\k)} \, ,  \label{eq:C2Phi_def} \\
C_{3,\Phi}(\x,\y)&= \int \frac{\d^3 k_1 \, \d^3 k_2 }{(2\pi)^6}\, e^{-i\k_1 \cdot \x - i \k_2 \cdot \y} \frac{B_\Phi(k_1,k_2,k_3)}{P_\Phi(k_1) P_\Phi(k_2) P_\Phi(k_3)}   \, , \label{equ:C3}
\end{align} 
with $\k_3 = - \k_1 - \k_2$.  
Crucially, as long as the integrand in (\ref{equ:C3}) is not an analytic function of any combination of the wavevectors, $C_{3,\Phi}(\x,\y \hskip 1pt)$ is a nontrivial function of two points. As a result, the probability distribution~(\ref{equ:219}) will depend on $\Phi(\x_i)$ at three separated points.

\vskip 4pt
Suppose that we measure the galaxy density contrast $\dgal(\x)$ which is a nonlinear function of $\delta$ and hence $\Phi$.  We further assume that we are still in the perturbative regime and that the relation between $\dgal$ and $\delta$ is invertible, so that 
\beq
\delta(\x) = \sum_n {\tilde b}_n \delta_g^n(\x)\,,
\label{equ:delta}
\eeq 
for some $\tilde b_n$.  Consider first the case of Gaussian initial conditions. Substituting (\ref{equ:delta}) into (\ref{equ:219}), with $f_{\rm NL} = 0$, gives \begin{align}
{\cal P}(\delta_g) =& \exp\bigg[ - \int \d^3 x_1 \d^3 x_2 \sum_{n,m} \tilde b_n \tilde b_m \delta_g^n(\x_1) \, C_2(\x_1-\x_2) \delta_g^m(\x_2) \bigg]\,,
\end{align}
where $C_2$ is the covariance of $\delta$, which is given by (\ref{eq:C2Phi_def}) with $P_\Phi(k)$ replaced by $P(k)$. 
Notice that all nonlinear terms that have been generated are still {\it bi-local}: they only depend on two points.  In contrast, the primordial non-Gaussian correlator is a function of three points.  In this sense, the information contained in the primordial three-point correlations is distinguishable from the secondary non-Gaussianity generated by nonlinear evolution,\footnote{In fact, this was already proven for the soft limits of inflationary correlators in~\cite{dePutter:2016moa}.  Unfortunately, in many models, including equilateral non-Gaussianity, constraints from the soft configurations alone are not competitive with the CMB constraints~\cite{Gleyzes:2016tdh}.} even if we do not know the coefficients $\tilde b_n$.  

\vskip 4pt
Figure~\ref{fig:causal} (on page~\pageref{fig:causal}) illustrates this pictorially.  Primordial equilateral non-Gaussianity implies nontrivial correlations for three points generated at the intersection of the past lightcones of these points.  The causal structure of the universe prevents such correlations between widely separated points from being generated at late times.  Nonlinear evolution at late times instead takes the primordial two-point correlations and modifies them locally by a nonlinear function.  This can generate nontrivial two-point correlations between composite operators (powers of $\delta$).  Higher-point correlations arise from the repeated application of the two-point statistics and the nonlinear map.

\vskip 4pt
This description makes our strategy clear: all of the information relevant to the nonlinear evolution is encoded in the two-point statistics of $\delta_g^n(\x)$.  However, in order to measure primordial non-Gaussianity, we must measure three-point correlations at separated points.  The fact that the nonlinear evolution will contribute to three-point statistics is not necessarily dangerous because these effects can be inferred from the measurement of the two-point statistics.

\vskip 4pt
So where did we go wrong in the previous section?  It is well known that the bispectrum is the optimal estimator for $\fnleq$~\cite{Heavens:1998jb,Babich:2005en,Creminelli:2005hu}, so how could the cosine between the bispectra not be an accurate description of the problem? The answer is that while the bispectrum is the optimal estimator for $\fnleq$, it is {\it not} the optimal estimator for all the $b_n$ coefficients.  There is more information about the coefficients $b_n$ in the map than that encoded in the galaxy power spectrum and bispectrum.\footnote{This same limitation applies to the modal decomposition of N-body simulations~\cite{Lazanu:2015rta, Lazanu:2015bqo}.  The bispectra of the simulations may be highly degenerate with the equilateral shape, but the maps aren't equivalent.}  Instead, the bi-local nature of the probability distribution manifests itself in terms of the higher-point statistics of $\dgal$.  However, because the biasing expansion is local, we do not need to including the full $N$-point functions, only the two-point statistics of the composite operators $\delta_g^n$.  If we compare the full maps and not just the low-point statistics, then we can capture all of the information encoded in these correlators and break the degeneracies with $\fnleq$.

\vskip 4pt
In principle, the results of the map-level analysis can also be recovered by the usual  perturbative analysis in Fourier space if we are careful to include a sufficiently high order of additional $N$-point correlators. For example, the degeneracy with the parameter $b_3$ would be broken by measuring the four-point function in addition to the bispectrum. What isn't manifest in this procedure, however, is the fact that locality dictates special relationships between the different correctors.  As we will see, because of locality the higher-order bias parameters---like~$b_3$---do {\it not} meaningfully affect the constraints on the primordial three-point function in the map-level analysis.  This means that these higher-order parameters are more strongly constrained by higher $N$-point functions---like the four-point function in the case of $b_3$---and therefore do not impact the measurement of primordial non-Gaussianity in the three-point function. This reduces the challenge of nonlinear biasing from an infinite set of contributions to just the leading quadratic operators.

\section{Map-Level Information}
\label{sec:map}

As explained in the previous section, at the level of the map, the nonlocal correlations created during inflation are distinguishable from the effect of local nonlinear evolution.  To make this more concrete, we will now discuss the likelihood for each realization of the map and show how it informs our knowledge of primordial non-Gaussianity.  Our analysis is closely related to~\cite{Schmittfull:2018yuk,Cabass:2019lqx,Cabass:2020nwf,Schmidt:2020viy,Schmittfull:2020trd}, but we will take a somewhat simplified approach that focuses on the local nature of the likelihood and its impact on the forecasts for $\sigma(\fnleq)$.

\vskip 4pt
First, we will consider the likelihood for a nonlinear map under the (strong) assumption that the initial realization of the Gaussian field is {\it known}.   In that case, there is {\it no} cosmic variance and our ability to measure primordial non-Gaussianity is only limited by the degeneracy with nonlinear evolution and by shot noise.  In this idealized context, it is easy to see that the degeneracy with nonlinear evolution is small.  This is a useful starting point as the salient features of this description will survive in our full forecasts.

\vskip 4pt
Next, we will extend our results to the case where the Gaussian initial conditions are {\it not known} and must therefore also be determined from the data.   
In the absence of biasing, we recover the same result for the $\fnleq$ measurement as from a standard bispectrum analysis.  However, we will see that the map-level likelihood includes vastly more information about the nonlinear evolution that helps to break the degeneracy with the primordial signal. 
 
\subsection{Without Cosmic Variance}

Suppose that we are given a specific realization of the Gaussian potential, $\Phi_{\rm G}(\x)$, and only the parameters $\fnleq$ and $b_n$ need to be determined by measuring the galaxy density contrast $\dgal$.  We write the non-Gaussian field as $\Phi(\k) = \Phi_{\rm G}(\k)+ \fnleq \Phi_{\rm NG}(\k)$,  with
\beq
\fnleq \Phi_{\rm NG}(\k) =\int \frac{\d^3 p \hskip 1pt \d^3 q}{(2\pi)^3}\frac{B_\Phi(\vec p,\vec q\hskip 1pt)}{6\, P_\Phi(p)P_\Phi(q) } \,\Phi_{\rm G}(\p\hskip 1pt) \Phi_{\rm G}(\q\hskip 1pt)\, (2\pi)^3 \delta_{\rm D}(\p+\q-\k) \, ,\label{equ:PhiNG}
\eeq
where $B_\Phi(\k_1,\k_2) \equiv B_\Phi(k_1,k_2,|\k_1+\k_2|)$ is the primordial bispectrum (see also~\cite{Schmidt:2010gw,Scoccimarro:2011pz}).
We define the model for the {\it linearly evolved} dark matter density contrast and the galaxies overdensities as before, but split the density contrast into a Gaussian and a non-Gaussian piece: 
\begin{align}
\delta_{\rm G}(\k) &= 
\T(k)\hskip 1pt \Phi_{\rm G}(\k) \,,\\
\delta_{\rm NG}(\k) &= 
\T(k)\hskip 1pt \Phi_{\rm NG}(\k) \, ,\label{eq:dNG}
\end{align}
where the linear transfer function $\T(k)$ was defined in (\ref{equ:deltaDEF}). 
To avoid clutter, we will drop the subscript on the Gaussian part, $\delta_{\rm G} \to \delta$.
Given a known realization of the linear field $\delta(\x)$, which we denote by $\bar \delta(\x)$, and an associated realization of the non-Gaussian term, $\bar \delta_{\rm NG}$, our model for the galaxy density field  is 
\begin{align}
\dgal(\x) &= \sum_{n=1}^\infty b_n [(\bar \delta+\fnleq \bar \delta_{\rm NG})^n](\x)\, \\
&\approx  \sum_{n=1}^\infty \Big(b_n [\bar \delta^n](\x) + n \fnleq b_n  [\bar\delta^{n-1} \,  \bar \delta_{\rm NG} ](\x)\Big) \, ,
\end{align}
where we used the small amplitude of the expected non-Gaussian signal to drop higher powers of $\bar \delta_{\rm NG}$ in the second line.  This model is related to the observed galaxy density by  
\beq\label{eq:dgalO_bar}
\dgalO(\x) = \sum_{n=1}^\infty \Big(\bar b_n [\bar \delta^n](\x) + n \bar{f}_{\rm NL}^{\rm eq}\bar b_n  [\bar \delta^{n-1} \,  \bar \delta_{\rm NG} ](\x)\Big) + \epsilon(\x)\,,
\eeq
where $\bar b_n$ are the fiducial values of the bias parameters.  The {\it stochastic bias}, $\epsilon(\x)$, defines how the formation of an object depends on physics beyond the long-wavelength density field.  Because $\epsilon(\x)$ is unknown, it is effectively a source of noise.  We will assume that it is a (unknown) realization of a Gaussian stochastic field with $\langle \epsilon(\x) \epsilon(\x^{\hskip 1pt \prime}) \rangle = N^2 \delta_{\rm D}(\x-\x^{\hskip 1pt \prime})$.  Given that galaxies are discrete objects and the density is continuous, the observed distribution of galaxies is never purely a function of $\delta$ alone.  To a good approximation~\cite{Cabass:2020nwf}, the difference is captured by shot noise, where $\epsilon(\x)$ is indeed Gaussian and $N^2 = 1/\bar n$, with $\bar n$ being the average density of galaxies.  We will see that the assumptions about the stochastic bias are largely unimportant when we measure modes with high signal-to-noise, $P(k) \gg N^2$.  
 
 \vskip 4pt
 The likelihood of the observed galaxy map $\dgalO(\x)$ is  
 \begin{align}
{\cal L} &= \exp \left( - \frac{1}{2 N^2} \int \d^3 x  \left(\dgal(\x) - \dgalO(\x) \right)^2 \right)  \label{equ:LnoCV}\\
&= \exp \Bigg( - \frac{1}{2 N^2} \int \d^3 x\,  \bigg(\sum_{n=1}^\infty \Big( b_n [\bar \delta^n](\x) + n \fnleq b_n  [\bar \delta^{n-1} \,  \bar \delta_{\rm NG} ](\x)\Big) - \dgalO(\x) \bigg)^2 \Bigg) \, , \label{equ:LnoCV2}
\end{align}
which is Gaussian because we have assumed that the only source of uncertainty, $\epsilon(\x)$, is a Gaussian random field.  We use this likelihood to define the Fisher matrix as
\beq
F_{i j} = - \frac{\partial^2}{\partial \theta_i \partial \theta_j} \log {\cal L} \Big|_{b_n =\bar b_n, \fnleq=0}\,, \label{eq:fisher}
\eeq
where $\theta_i \in (\{ b_n \}, \fnleq)$ are the model parameters and we are assuming a fiducial value of $\bar{f}_{\rm NG}^{\rm eq}= 0$.  Taking derivatives of (\ref{equ:LnoCV2}), we find 
\beq
\label{eq:fisher_NoCV} 
\begin{aligned}
F_{n,m} &= \int \d^3 x \,\frac{[\bar \delta^{n}](\x) [\bar \delta^m](\x)}{N^2} \,, \\ 
F_{n,{\rm eq}}  &= \sum_m \int \d^3 x\, m \bar b_m \frac{[\bar \delta^{n}](\x)[\bar \delta^{m-1}\bar \delta_{\rm NG}](\x)}{N^2} \,,\\ 
F_{\rm eq,eq} &= \int \d^3 x \, \bar b_1^2 \frac{[\bar \delta_{\rm NG}]^2(\x)}{N^2}\, ,
\end{aligned}
\eeq
where we have dropped subleading $\bar \delta^{m-1} \bar \delta_{\rm NG}$ terms in $F_{\rm eq, eq}$. 

\vskip 4pt
We use the ergodic theorem to replace the spatial averages in the Fisher matrix elements with statistical averages over the distribution (times the volume of the survey $V$).  At this point, we see the benefit of working with renormalized operators, where $\langle [\delta^n](\x) [\delta^m](\x^{\hskip 1pt \prime}) \rangle' =0$ for $m\neq n$.  This simplifies the off-diagonal contributions to the Fisher matrix, so that
\begin{align}
F_{n,m} & = V \delta_{n,m} \int \d^3 x\, \frac{\langle [ \delta^{n}](\x)[ \delta^{n}](\x)\rangle'}{N^2} \ ,\\
F_{n,\rm eq} &= V \int \d^3 x\, (n-1) \bar b_{n-1}\, \frac{\langle [ \delta^{n}](\x)[ \delta^{n-2}  \delta_{\rm NG}](\x)\rangle'}{N^2} \, ,\\
F_{\rm eq,eq} &=V \int \d^3 x \, \bar b_1^2\, \frac{\langle[\bar \delta_{\rm NG}]^2(\x) \rangle'}{N^2}\, .
\end{align}
The essential feature of note in these expressions is that $F_{n,n} \propto \delta^{2n}$, $F_{n, {\rm eq}} \propto  \delta^{2n}$ and $F_{\rm eq,eq} \propto \delta^4$.  This will mean that higher-order nonlinearities aren't degenerate from the measurement of $\fnleq$.  Concretely, given these expressions for the Fisher information, we can define a cosine on the space of maps by direct analogy with the bispectrum cosine in (\ref{eq:cosine_B}):
\beq
\cos([\delta^n] , \delta_{\rm NG}) = \frac{F_{n, \rm eq} }{\sqrt{F_{n , n} F_{\rm eq, eq} }}  \propto \left( \langle \delta^2(\x) \rangle'\right)^{(n-2)/2}\, ,
\label{equ:316}
\eeq
where the final equality shows the scaling with powers of $\delta(\x)$.
 This cosine quantifies how distinguishable the non-Gaussian contributions from $[\delta^n]$ and $\delta_{\rm NG}$ are in the map $\dgalO$. We see that, at the map level, the $n > 2$ contributions are {\it not} degenerate with the primordial non-Gaussianity as long as we are in the perturbative regime where $\delta < 1$.  This essential feature of the map-level approach will persist in the more realistic case with cosmic variance.

\subsection{With Cosmic Variance}
\label{sec:Section3-2}

Unfortunately, we do not know the initial Gaussian map of the universe and therefore we must isolate any primordial non-Gaussianity from late-time nonlinearities and from random statistical fluctuations of a Gaussian field.  
Our goal therefore is to remove our knowledge of the initial conditions from the likelihood while still working at the level of the galaxy map.  We will  show that, in the absence of nonlinear biasing (i.e.~holding $b_{n>1} =0$ fixed), we reproduce the Cramer--Rao bound for the measurement of $\fnleq$, and that the additional information in the map beyond the bispectrum breaks the degeneracies for $b_{n>2}\ne 0$.

\vskip 4pt
To account for the fact that the Gaussian potential $\Phi_{\rm G}$---and hence the linearly evolved (Gaussian) density contrast $\delta(\k) \equiv {\cal T}(\k) \Phi_{\rm G}(\k)$---is unknown, the likelihood function~(\ref{equ:LnoCV}) must be corrected by the likelihood for $\delta$; cf.~(\ref{equ:219}). 
The modified likelihood function is
\begin{align}
\label{eq:like_map}
{\cal L} &= \exp \left(- \frac{1}{2 N^2} \int \d^3 x\,  \big(\dgal(\x) - \dgalO(\x) \big)^2  -\int \d^3 x \hskip 1pt \d^3 x' \, \Phi_{\rm G}(\x)\hskip 1pt  C_{2,\Phi}(\x-\x^{\hskip 1pt \prime})\hskip 1pt  \Phi_{\rm G}(\x^{\hskip 1pt \prime})  \right) \\
&\approx
\exp \left( -\int \frac{\d^3 k }{(2\pi)^3} \left[ \frac{1}{2 N^2} \left|\dgal(\k) - \dgalO(\k) \right|^2+ \frac{|\delta(\k)|^2}{2 P(k)}   \right] \right) ,
\end{align}
where we can assume that $P(k)$ is known, as it is very accurately measured by the CMB. 

\vskip 4pt
From a computational point of view, one could analyze the maps directly using the likelihood to determine the best fit for the parameters $b_n$, $\fnleq$ and the realization of the initial Gaussian map~$\bar \delta$. One could use~(\ref{eq:like_map}) as the likelihood for the observed map of galaxies (suitably generalized to redshift space~\cite{Cabass:2020jqo}) and solve the high-dimensional minimization problem by brute force.  Of course, good approximate methods exist for these types of problems (using machine learning~\cite{Taylor:2019mgj,Dai:2020ekz,Modi:2020dyb,Modi:2021acq,Makinen:2021nly,Hassan:2021ymv,Villaescusa-Navarro:2021pkb,Villaescusa-Navarro:2021cni}) and this approach is potentially the best way to perform a map-level analysis in practice.  However, as our interest is in understanding the nature of the cosmic information in the maps, we will take the following analytic approach:  we will first guess a model of the initial Gaussian map, $\bar \delta_{\rm guess}(\x)$, and find the maximum likelihood values of the parameters $b_n$ and $\fnleq$ for the observed map, $\dgalO(\x)$, while holding $\bar \delta_{\rm guess}(\x)$ fixed. Then, holding $b_n$ and $\fnleq$ fixed, we  want to find the Gaussian map that maximizes the likelihood, $\dbarO(\x)$, in terms of the observed map, $\dgalO(\x)$.  We can repeat this procedure iteratively until we reach the maximum likelihood point for all the parameters and the galaxy map. 

\vskip 4pt
As the first step in this iterative procedure, we find the maximum likelihood values of $\fnleq$ and $b_n$, when $\bar b_n$ and $\bar f_{\rm NL}^{\rm eq}$  are the fiducial values.  This maximization is presented in Appendix~\ref{app:likelihoods} and, for high signal-to-noise, it is easy to see that the maximum likelihood parameters agree with the fiducial values $b_n = \bar b_n$ and $\fnleq = \bar f_{\rm NL}^{\rm eq}$.

\vskip 4pt
Next, we want to determine the realization of the linearly evolved field, $\bar \delta(\x)$, from the observed galaxy map, $\dgalO(\x)$, while holding $b_n = \bar b_n$ and $\fnleq = \bar f_{\rm NL}^{\rm eq}$ fixed. A priori, one might imagine that inverting this map is difficult.  After all, if we could easily determine the exact realization of the density field, $\bar \delta(\x)$, then we could measure the primordial non-Gaussianity directly in the initial conditions without any nonlinearity.  However, if we {\it knew} the bias parameters, $b_n$, then the inversion is not difficult---at least at high signal-to-noise where $\epsilon(\x)\to 0$. Specifically, for $\fnleq=0$ and $\epsilon(\x)=0$, we make an educated guess for $\bar \delta(\x)$ inspired by~(\ref{equ:delta}):
\begin{align}\label{eq:delta_guess}
\bar \delta_{\rm guess}(\x) &\ =\ \sum_{n} \tilde b_n  [(\dgalO)^n] (\x)\nonumber \\[8pt]
 &\ \to\ b_1^{-1} \Big(\bar b_1 \bar \delta(\x) +\bar b_2 [\bar \delta^2](\x) + \cdots \Big) -b_2 b_1^{-3}\, \bar b_1^2 [ \bar \delta^2](\x) + \cdots \nonumber \\[8pt]
 &\ =\ \bar \delta(\x) + {\cal O}(\bar \delta^3) \, .
\end{align}
We see that, for fixed $b_n =\bar b_n$, there is a natural inversion of the map from $\dgalO \to \bar \delta$, given by $\tilde b_1 = b_1^{-1}$, $\tilde b_2 = - b_2 b_1^{-3}$, etc.  This procedure works order by order: Given the error in the inversion at order $(\dgalO)^n$, we remove the leading-order error with an appropriate choice of coefficient at order $(\dgalO)^{n+1}$;  this can be continued until we reach the desired level of precision.  The challenge with inverting the observed map is the {\it uncertainty} in $b_n$ which propagates into an uncertainty in $\bar \delta$.  Our imperfect knowledge of the inversion leads to an imperfect knowledge of the initial conditions. This interplay between $b_n$, $\fnleq$ and $\bar \delta$ will lead to the map-level Fisher matrices.  

\vskip 4pt
In practice, we do not want to simply guess a solution that returns $\bar \delta(\x)$ in the idealized limit.  Instead, we want to determine the {\it maximum likelihood map}, $\dbarO(\x)$, given the data $\dgalO(\x)$. Our observed map $\dgalO(\x)$ also contains noise in the form of the stochastic bias parameter $\epsilon(\x)$ whose impact we wish to minimize.  We therefore need a parameterization of this inversion that allows us to filter out the noisiest modes.  Inspired by~(\ref{eq:delta_guess}), we make the following ansatz
\beq
\dbarO(\x) \ =\  \sum_{n} \tilde b_n \star [(\dgalO)^n] (\x) \ -\ \fnleq  \bar \delta_{\rm NG}[\dgalO](\x) \, ,  \label{equ:Inversion}
\eeq
where the constant coefficients $\tilde b_n$ have been replaced with {\it filter functions}, $\tilde b_n(\x)$, and
$f \star g(\x) \equiv \int \d^3 x' f(\x^{\hskip 1pt \prime}) g(\x-\x^{\hskip 1pt \prime})$ denotes a convolution.
The function $\bar \delta_{\rm NG}$ removes the primordial non-Gaussianity from the map.  Note that (\ref{equ:Inversion}) reduces to~(\ref{eq:delta_guess}) when $\fnleq = 0$ and $\tilde b_n(\x) = \tilde b_n \delta_{\rm D}(\x)$.  By definition, $\dbarO(\x)$ will be our best estimate for the Gaussian initial conditions and therefore the model for the galaxy density is determined by these initial conditions,
\begin{align} \label{eq:gal_inv}
\delta_g(\x) &\ =\  \sum_n b_n  \left[\left(\dbarO(\x)+ \fnleq \delta_{\rm NG}[\dbarO](\x) \right)^n \right] ,
\end{align}
where $\delta_{\rm NG}$ is given by~(\ref{eq:dNG}), evaluated with the maximum likelihood solution for $\dbarO(\x)$ in place of $\delta(\x)$.  

\vskip 4pt
Our task is to find the specific functions $\tilde b_n(\x)$ that give the maximum likelihood solution for $\dbarO(\x)$.  It will be easier to solve this in Fourier space, so that the convolutions become multiplications by $\tilde b_n(\k)$.  The details are given in Appendix~\ref{app:likelihoods}. Up to second order in $\dgalO$, and at high signal-to-noise $P(k) \gg N^2$, we find
\begin{align}
 \tilde b_1(\k) &=  \frac{b_1 P(k)}{b_1^2 P(k)+N^2} \, , \\
 \tilde b_2 (\k) &=  - \frac{b_1 P(k)}{b_1^2 P(k)+N^2} \, \frac{b_2}{b_1^{2}} \, ,\\
 \bar \delta_{\rm NG}(\k) &= \frac{b_1^2 P(k)}{b_1^2 P(k)+N^2} \,\delta_{\rm NG}[\tilde b_1\star \dgalO](\k)\, .
 \label{equ:324}
 \end{align}
This result makes intuitive sense: in the limit of high signal-to-noise, $N^2 \ll P(k)$, it reproduces our guess in~(\ref{eq:delta_guess}), including the obvious generalization to $\fnleq \neq 0$.  It is not hard to see that this pattern continues at ${\cal O}((\dgalO)^n)$, such that $\tilde b_n(k) \supset -b_n b_1^{-n+1} P(k)/(b_1^2P(k)+N^2)$.  

\vskip 4pt
For simplicity, we will now set $b_1 \equiv 1$. The only nontrivial impact of $b_1$ is to change the effective shot noise $N^2 = 1/\bar n\to 1/(b_1^2 \bar n)$.  Otherwise, we can include it by a simple rescaling of the coefficients. Equation (\ref{equ:324}) then becomes
\beq
 \bar \delta_{\rm NG}(\k) = \frac{P(k)}{P(k) + N^2}  \int \frac{\d^3 q\hskip 1pt  \d^3 p}{(2\pi)^3} \frac{B_{\rm eq}(\p, \q\hskip 1pt)}{(P(p)+N^2)(P(q)+N^2) } \, \dgalO(\p\hskip 1pt) \dgalO(\q\hskip 1pt) \, \delta_{\rm D}(\p+\q-\k) \, , 
\eeq
where $B_{\rm eq}(\p, \q\hskip 1pt)$ is the linearly evolved primordial bispectrum.  Going beyond quadratic order in~$\dgalO$, it is convenient to work with renormalized operators.  However, to ensure that both~(\ref{equ:Inversion}) and~(\ref{eq:gal_inv}) are expanded in terms of renormalized operators, we need to shift the definition of $\tilde b_1(\k)$. In the limit $P(k) \gg N^2$, we get
\beq
\begin{aligned}
 \tilde b_1(\k) \to 1 + 4 b_2^2 \sigma^2 & + \frac{2}{3} \fnleq b_2\int \frac{\d^3 p}{(2\pi)^3} \frac{B_{\rm eq}(\p,\k-\p\hskip 1pt)}{P(|\k-\p\hskip 1pt |)}    \\
&+ \frac{1}{9} (\fnleq)^2 \int \frac{\d^3 p}{(2\pi)^3} \frac{B_{\rm eq}(\p,\k-\p\hskip 1pt) B_{\rm eq}(-\p, \k)}{P(|\k-\p\hskip 1pt |) P(p) P(k)}   \, ,
\end{aligned}
\eeq
where $\sigma^2  \equiv \int \d^3 k \,P(k)/(2\pi)^3$.  See Appendix~\ref{app:beyond_linear} for details. 

\vskip 4pt
At low signal-to-noise, $N^2 \gtrsim P(k)$, the maximum likelihood maps become significantly more complicated.  However, as we saw in Section~\ref{sec:challenge}, the forecasts we want to compare to are limited by theoretical errors and not shot noise.  For the purposes of the present discussion, it is therefore sufficient to keep only the leading terms in the $P(k) / N^2$ expansion.  Furthermore, a proper map-based analysis would find the maximum likelihood maps numerically, thus circumventing the need for complicated analytic expressions.  

\vskip 4pt
With the maximum likelihood maps in hand, we can now expand the likelihood to second order in $b_n$ and $\fnleq$ around the maximum likelihood point and determine the Fisher matrix.  Again,  the algebraic details can be found in Appendix~\ref{app:likelihoods}. 
The resulting Fisher matrix  is very similar to~(\ref{eq:fisher_NoCV}), but with $N^2 \to P(k) +N^2$ and $\delta \to \dgalO$:  
\beq\label{equ:FisherWith_Obs}
\begin{aligned}
F_{n,m} &=  \int  \frac{\d^3 k}{(2\pi)^3}  \frac{[(\dgalO)^n](\k) [(\dgalO)^m](-\k) + n \delta_{n,m} n! \sigma^{2n-2} (2|\dgalO(\k)|^2 - \sigma^2)}{P(k) + N^2} \,,\\[4pt]
F_{n, \rm eq} &=   \sum_m \int  \frac{\d^3 k}{(2\pi)^3} \, m \bar b_m \frac{[(\dgalO)^n](\k)[(\dgalO)^{m-1}\delta_{\rm NG}[\dgalO] ](-\k)}{P(k)+N^2}   \\
& \quad +  \frac{2}{3} \delta_{n,2} \int \frac{\d^3 p}{(2\pi)^3} \frac{B_{\rm eq}(\p,\k-\p\hskip 1pt)}{(P(|\k-\p\hskip 1pt |)+N^2) (P(k)+N^2)^2} (2|\dgalO(\k)|^2 - \sigma^2)\, , \\[6pt]
F_{\rm eq, eq} & = \int  \frac{\d^3 k}{(2\pi)^3}  \frac{\delta_{\rm NG}[\dgalO](\k) \hskip 2pt \delta_{\rm NG}[\dgalO](-\k)}{P(k)+N^2} \\
& \quad +  \frac{1}{9} \int \frac{\d^3 p}{(2\pi)^3} \frac{B_{\rm eq}(\p,\k-\p\hskip 1pt) B_{\rm eq}(-\p, \k)}{(P(|\k-\p\hskip 1pt |)+N^2) (P(p)+N^2) (P(k)+N^2)^2} (2|\dgalO(\k)|^2 - \sigma^2)\, .
\end{aligned}
\eeq
Note that this Fisher matrix includes only the leading contributions to the maximum likelihood map, $\dbarO(\x)$.  As we have seen, the higher-order terms in $\dbarO(\x)$ eliminate the nonlinearities of $\dgalO(\x)$ from the maximum likelihood map of the linearly-evolved field, $\dgalO(\x)$.  We therefore anticipate that at higher orders, our procedure will converge to $\dbarO(\x) \approx  \bar \delta(\x)$.   Working at large (but finite) signal-to-noise, one finds residual factors of $P(k) +N^2$ in the Fisher matrices, so that 
\beq\label{equ:FisherWith}
\begin{aligned}
F_{n,m} &\to V  \int   \frac{\d^3 k}{(2\pi)^3}  \frac{  \langle [\delta^n](\k) [\delta^m](-\k) \rangle'+ n \delta_{n,m} n! \sigma^{2n}}{P(k) + N^2} \,,\\[6pt]
F_{n, \rm eq} &\to  V \sum_m \int  \frac{\d^3 k}{(2\pi)^3} \, m \bar b_m \frac{ \langle [\delta^n](\k)[\delta^{m-1}\delta_{\rm NG}](-\k)\rangle'}{P(k)+N^2}   \\
& \quad \ \  + \frac{2}{3}\delta_{n,2} \int \frac{\d^3 p}{(2\pi)^3} \frac{B_{\rm eq}(\p,\k-\p\hskip 1pt)}{(P(|\k-\p\hskip 1pt |)+N^2) (P(k)+N^2)} \, , \\[6pt]  
F_{\rm eq, eq} & \to V \int  \frac{\d^3 k}{(2\pi)^3}  \frac{\langle \delta_{\rm NG}(\k) \delta_{\rm NG}(-\k) \rangle' }{P(k)+N^2}\\
&\quad \ \ +  \frac{V}{9} \int \frac{\d^3 p}{(2\pi)^3} \frac{B_{\rm eq}(\p,\k-\p\hskip 1pt ) B_{\rm eq}(-\p, \k)}{(P(|\k-\p\hskip 1pt |)+N^2) (P(p)+N^2) (P(k)+N^2)} \, ,
\end{aligned}
\eeq
where we again replaced spatial averages with statistical averages over the distribution of~$\delta(\x)$. The inverse powers of $P(k) +N^2$, rather than simply $P(k$), are meaningful as they show our maximum likelihood formulas reproduce the expected suppressions in the presence of noise that are not manifest in the likelihood~(\ref{eq:like_map}).

\vskip 4pt
For $n= 2$, we get 
\begin{align}
F_{2,2} & =  \frac{2V}{3}  \int  \frac{\d^3 p \hskip 1pt \d^3 q}{(2\pi)^6}\,  \frac{ \left(P(|\p+\q\hskip 1pt|)+P(p)+P(q)\right)^2}{(P(|\p+\q\hskip 1pt |)+N^2)(P(p)+N^2) (P(q)+N^2) } \,, \\[6pt]
F_{2, \rm eq}  &=  \frac{V}{3} \int  \frac{\d^3 p \hskip 1pt \d^3 q}{(2\pi)^6}\,  \frac{\left(P(|\p+\q\hskip 1pt |) +P(p)+P(q)\right) B_{\rm eq}(\p, \q\hskip 1pt)}{(P(|\p+\q\hskip 1pt |)+N^2)(P(p)+N^2) (P(q)+N^2)} \,, \\[6pt]
F_{\rm eq,eq} &=  \frac{V}{6} \int  \frac{\d^3 p \hskip 1pt \d^3 q}{(2\pi)^6}\,  \frac{B_{\rm eq}^2(\p, \q\hskip 1pt)}{(P(|\p+\q\hskip 1pt|)+N^2)(P(p)+N^2) (P(q)+N^2)} \, ,\label{eq:fnl_wCV}
\end{align}
which is of the same form as the Fisher information of the optimal bispectrum estimator\footnote{This statement is true at leading order in $\delta$, assuming the same $k_{\rm max}$ for all correlators.  At higher orders, there is also information relevant to $b_2$ encoded in the power spectrum.  In perturbative analyses like~\cite{Oddo:2021iwq,Ivanov:2021kcd}, one is able to use a larger $k_{\rm max}$ in the power spectrum than in the bispectrum, which can compensate for the $\delta$-suppression. Given that one cannot achieve competitive constraints on $\fnleq$ from the power spectrum alone~\cite{Gleyzes:2016tdh}, this additional information will, at best, reduce the degeneracy between $\fnleq$ and $b_2$. } that defined our cosine in~(\ref{eq:bispectrum_dot}), with $P \to P+N^2$ to account for shot noise.  To make the result recognizable, we have restored factors of $P(k)/(P(k)+N^2) \to 1$ that we eliminated in the high signal-to-noise regime.  In addition, as shown in Section~\ref{app:map_bi}, the formulas can be arranged in a more permutation symmetric form that makes the equivalence with the bispectrum analysis manifest.  

\vskip 4pt 
For $n > 2$, the map-level Fisher results will differ from those using the bispectrum alone.  Importantly, however, the result in~(\ref{equ:FisherWith}) shows the same suppression of the higher-order terms that we observed in the absence of cosmic variance. In particular, the diagonal terms scale as $F_{n,n} = {\cal O}(\delta^{2n})$ and $F_{\rm eq,eq} = {\cal O}(\delta^{4})$, while the leading off-diagonal terms are suppressed, $F_{n, \rm eq} = {\cal O}(\delta^{2n}) \ll \sqrt{F_{\rm eq, eq} F_{n,n}} = {\cal O}(\delta^{n+2})$.  Using the map-level cosines in~(\ref{equ:316}), we therefore conclude that the degeneracy between $b_n$ and $\fnleq$ is small for $n>2$.  In contrast, the cosines defined in~(\ref{eq:cosine_B}) in terms of the bispectra alone are independent of the amplitudes of the individual bispectra and thus don't experience the same suppression.   The fundamental reason for the suppression of the map-level cosine is that the operators  $[\delta^n]$ introduce a nonzero $(n+1)$-point function at leading order in $\delta$. The off-diagonal terms $F_{n, \rm eq}$ then no longer capture the comparison of two bispectrum shapes, but instead are associated with the comparison between an $n$-point function and a bispectrum, which would be completely independent if it were not for higher-order effects in $\delta$.

\begin{figure}[t!]
\centering
\includegraphics[scale=1.]{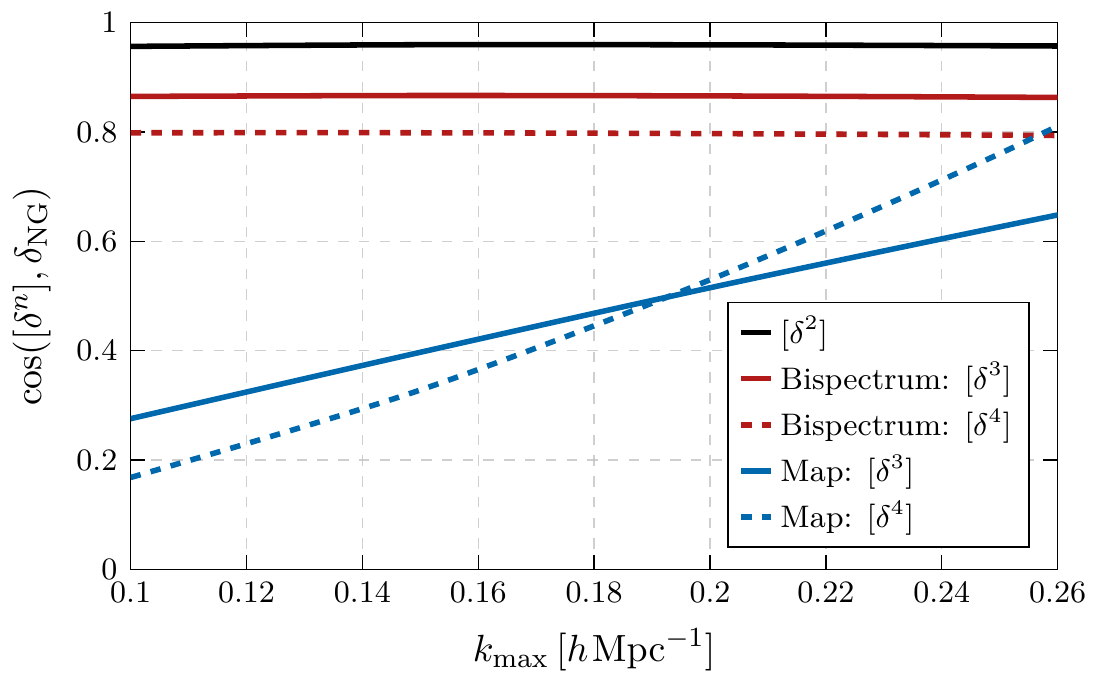}
\caption{Cosines of $[\delta^2]$, $[\delta^3]$, $[\delta^4]$ with $\delta_{\rm NG}$, defined at the map level in (\ref{equ:MapCosine}) and for the bispectrum in~(\ref{eq:cosine_B}).  For $[\delta^2]$, these two definitions are equivalent. The map level cosines are calculated using the forecasting methods described in Section~\ref{sec:forecasts}. We see that the map-level information dramatically reduces the cosine (degeneracy) when $k_{\rm max} < \knl \approx 0.25 \,  h \, {\rm Mpc}^{-1}$. }
\label{fig:cosines}
\end{figure}

\vskip 4pt
Let us see more explicitly that, at the map level, the contributions from $b_{n>2}$ are essentially orthogonal to $\fnleq$.  We again assume a power-law spectrum, as in~(\ref{eq:power_law}), to determine the scaling behavior with $k_{\rm max}$ and $\knl$.
At leading order in $\delta$, this gives  
\beq
\begin{aligned}\label{eq:scaling_fisher}
F_{n,m} &\propto V \delta_{n,m} \left( \frac{k_{\rm max}}{\knl} \right)^{(\Delta+3) (n-1)} \,,\\
F_{n,\rm eq } &\propto V \Delta_\Phi^2 \left( \frac{k_{\rm max}}{\knl}\right)^{(\Delta+3)(n-1)}\,, \\
F_{\rm eq, eq} &\propto V  \Delta_\Phi^4 \left( \frac{k_{\rm max}}{\knl}\right)^{\Delta+3}\,,
\end{aligned}
\eeq
and the cosines are 
\beq
\cos([\delta^n], \delta_{\rm NG}) = \frac{F_{n,{\rm eq}}}{\sqrt{F_{n,n} F_{\rm eq, eq}}} \propto \left(\frac{k_{\rm max}}{\knl} \right)^{(\Delta+3)(n-2)/2}\,.
\label{equ:MapCosine}
\eeq
Taking $k_{\rm max} < \knl$, so that we are still in the perturbative regime,  the cosine is power law suppressed for all $n> 2$.  While there is a significant degeneracy between $b_2$ and $\fnleq$, all higher-order nonlinearities are increasingly orthogonal to $\fnleq$ and thus do {\it not} pose a serious challenge to measuring $\fnleq$, when the map-level information is included.  
Figure~\ref{fig:cosines} shows the behavior of the cosines for the true $\Lambda$CDM spectrum (not just the power law ansatz).

\vskip 4pt
It is worth highlighting that the benefit of the map-level analysis does {\it not} just come from the fact that modes are more linear at small $k$.  If we set the transfer function $\T(k) \to 1$, then all Fisher matrix elements are scale invariant with $\Delta =-3$, so that the resulting Fisher matrices have $F_{n,n} \propto \Delta_\Phi^n$,  $F_{n, \rm eq} \propto \Delta_\Phi^{n+2}$ and $F_{\rm eq,eq} \propto \Delta_{\Phi}^{4}$. In this case, the cosines are still highly supressed, $\cos([\delta^n], \delta_{\rm NG}) \propto \Delta_\Phi^{(n-2)/2}$, for $n>2$.  In this precise sense, the map-level cosines are suppressed by the amount of nonlinearity at the scale at which we are getting our information.  In the scaling universes of interest and in $\Lambda$CDM, most information in the map comes from $k_{\rm max}$.  In a scale-invariant universe with $b_{n} = O(1)$, $\Delta_\phi$ sets the amplitude of nonlinearity on all scales and controls the size of the map-level cosines.

\begin{center}
***
\end{center}

\noindent
Before moving on, let us summarize the main result and explain why the map-level analysis is more powerful that a conventional bispectrum analysis.  As expected, the Cramer--Rao bound from the map-level analysis and from a bispectrum analysis are essentially identical.  Said differently, it is not possible to measure $\fnleq$ more precisely than with an optimal bispectrum estimator defined in~\cite{Creminelli:2005hu}.  However, nonlinear corrections of the form $\delta^n$ do more than just change the bispectrum; the dominant contribution to its Fisher matrix is from a two-point function of composite operators, $\langle [\delta^n](\x) [\delta^n](\x^{\hskip 1pt\prime}) \rangle$.  Written in terms of Fourier modes, $\delta(\k)$, this information would take the form of a $2n$-point function with a very specific analytic structure in $\k$.  One might have worried that introducing additional higher-point information would be computationally expensive, but the fact that the relevant information reduces to two-point functions in position space suggests that it is a computationally tractable problem (see also~\cite{Munchmeyer:2019wlh} for a related example).

\section{Forecasted Sensitivities} \label{sec:forecasts}

In the previous section, we demonstrated how the map-level information is, in principle, sufficient to distinguish between primordial non-Gaussianity and late-time nonlinearities.  These results were largely analytic and based on scaling behavior.  
Now, we would like to calculate the Fisher matrix more explicitly, replacing the scaling ansatz with the true power spectrum in $\Lambda$CDM. Our forecasts will assume a fixed volume survey at $z=0$, so that the nonlinear scale is fixed for all modes in the survey.  As a consequence, our forecasts will necessarily be more pessimistic (at both the map and bispectrum level) than a realistic survey that covers a range of redshifts.  Our emphasis is instead on the relative impact of nonlinearity at the map and bispectrum level.  We will explain in the end how these results inform the potential reach of planned surveys.  

\vskip 4pt
The goal is this section is to show explicitly that the map-level $\fnleq$ constraints do not suffer from the large degeneracies with the higher-order coefficients of the biasing expansion that arise in a bispectrum-only analysis.   We will compare the volume scalings of the map-level and bispectrum forecasts with theoretical errors and show that the map-level results recover the naive volume scaling, even when marginalizing over additional bias parameters.  This means that bispectrum forecasts with theoretical errors overestimate the fundamental limitations in measuring $\fnleq$ presented by higher-order nonlinearity or unknown short-distance physics.

 \vskip 4pt
Our modeling of nonlinearities will be restricted to the bias expansion given in~(\ref{equ:biasing}).  This expansion was shown in~\cite{Schmittfull:2018yuk} to reproduce the halos produced in simulations, up to stochastic bias, and thus does not represent a significant simplification of the physics of structure formation.  However, in writing this bias expansion, we do neglect the effect of bulk flows, which were included in~\cite{Schmittfull:2018yuk} by ``shifting" the operators according to the Zel'dovich approximation.  In principle, these bulk flows can be removed by ``reconstruction"~\cite{Eisenstein:2006nk} and thus we are only neglecting the incompleteness of reconstruction and the associated noise.  In addition, observations are subject to redshift space distortions~\cite{Percival:2008sh}, whose inclusion lies beyond the scope of this work.  Recent work~\cite{Ivanov:2021haa} suggests that the impact of redshift space distortions can be mitigated in power spectrum analyses and we therefore do not expect this to be a fundamental limitation at the map level either.

\subsection{Bispectrum Forecasts}

As a point of reference, we will briefly review the standard bispectrum analysis of a
galaxy survey.  We define an estimator for every bispectrum template $B_i$ that we want to constrain.  The exact estimator here is unimportant, the only feature we need is that it is a sum over three factors of the Fourier modes $\dgalO(\k)$ weighted by a template.  These estimators return measurements of the parameters $\fnleq$, $b_2$, $b_{{\cal G}_2}$, etc.  The variance of the bispectrum is set by the six-point function, which in the Gaussian limit is given by three factors of the power spectrum $P(k)$.  Putting all of this together, the Fisher matrix for two bispectra $B_i$ and $B_j$ is~\cite{Smith:2006ud}
\beq\label{eq:bispectrum_fisher}
\hspace{-0.05cm}F_{ij} = \frac{V}{6} \int  \frac{\d^3 k_1}{(2\pi)^3}   \frac{1}{k_1}\frac{3!}{(2\pi)^2}  \int\limits_{k_1/2}^{k_1} k_2\hskip 1pt \d k_2 \int\limits_{k_1-k_2}^{k_2} \hspace{-0.15cm}k_3 \hskip 1pt \d k_3  \frac{B_i(k_1,k_2, k_3) B_j(k_1,k_2, k_3)}{\left(P(k_1) +N^2 \right) \left(P(k_2) +N^2 \right) \left(P(k_3) +N^2 \right)} \,,
\eeq
where $N^2 \equiv 1/\bar n$ is the shot noise.  

\vskip 10pt
\noindent
{\bf Quadratic biasing}---Our baseline bispectrum forecasts will assume quadratic biasing, with the associated biasing parameters $b_2$ and $b_{\rm {\cal G}_2}$. The absence of higher-order biasing parameters, like $b_3$ and $b_4$, will be included as theoretical errors~\cite{Baldauf:2016sjb}. 
In the previous section, we saw that the map-level and bispectrum forecasts are equivalent for the parameters $b_2$, $b_{\rm {\cal G}_2}$, $\fnleq$, because the dominant effect on the map is the induced bispectrum.  This means that one can think of the Fisher matrix contributions for $b_2$, $b_{\mathcal{G}_{2}}$, $\fnleq$ as either map-level or bispectrum forecasts.  In addition, we will also include the leading nonlocal term, $b_{\partial^2} R_*^2 \partial^2 (b_1 \delta + b_2 \delta^2)$, to diagnose the role of locality in the constraining power. We will take $R^{-1}_*= 0.3 \, h \, {\rm Mpc}^{-1}$, although that choice will not impact the forecasts as it can be absorbed into the fiducial value of $b_{\partial^2}$.
The relevant bispectra of the biasing model then are
\begin{align}
 B_{2}  &=  2  \left[  P (k_1) P (k_2) +P (k_1) P (k_3)+  P (k_2) P (k_3)\right] , \label{equ:B2} \\[6pt]
B_{{\mathcal{G}_{2}}}  &= 2 \left[ \left( \mu_{12}^2- 1\right) P (k_1) P (k_2)  + {\rm perms} \right] , \\[6pt]
B_{\partial^2}  &=  - R_*^2 (k_1^2+k_2^2+k_3^2)\, B_{2} (k_1,k_2,k_3) \,,
\end{align}
where $\mu_{12}^2 \equiv \vec k_1 \cdot \vec k_2/(k_1k_2)$. The inflationary bispectrum $B_{\rm eq}$ was defined in (\ref{equ:Beq}).
Substituting these bispectra into the Fisher matrix~(\ref{eq:bispectrum_fisher}) then allows us to forecast the constraints on $b_2$, $b_{\mathcal{G}_{2}}$, $ b_{\partial^2}$ and $\fnleq$, choosing $\bar b_{\cal O} =1$ and $\bar{f}_{\rm NL}^{\rm eq} =0$
as the fiducial values.  

\vskip 10pt
\noindent
{\bf Theoretical errors}---In~\cite{Baldauf:2016sjb}, it was proposed that theoretical errors should be included in the bispectrum forecasts by changing the Fisher matrix to 
\beq\label{eq:bispectrum_TE}
F_{ij} \to \int  \sum_{T,T'} B_i(T) B_j(T^\prime) C_{T T^{\prime}}\,,
\eeq
where $T$ and $T'$ are the configurations of $\k_1, \k_2, \k_3$ forming a closed triangle and 
\beq
C^{-1}_{T T^{\prime}}=\frac{(2 \pi)^{3}}{V(z_{i})} \frac{f_{\text {sky }}^{-1}}{\d k_{1} \d k_{2} \d k_{3}} \frac{\prod_{i=1}^{3}\left(P\left(k_{i}\right)+N^2\right)}{k_{1} k_{2} k_{3}} \,\delta_{T T^{\prime}}+ (C_{e,B})_{T T^{\prime}} \, .
\eeq
We refer to $(C_{e,B})_{T T^{\prime}} =E^2_{B}(k_1,k_2,k_3)  \,\delta_{T T^{\prime}} $ as the {\it (bispectrum) theoretical error}, to be distinguished from the power spectrum theoretical errors introduced in~(\ref{eq:C_e}). We will mostly be concerned with the errors that arise when cubic and quartic operators are omitted from the biasing model.
In~\cite{Baldauf:2016sjb}, these errors were estimated as 
\beq
E_{B}(k_{1}, k_{2}, k_{3})= 3B_{2}(k_{1}, k_{2}, k_{3})\, 
 \left\{\begin{array}{ll}
\big( \hat k_t / 0.31\big)^{1.8} &  \qquad b_3 \text{ error}\,,\\[6pt]
\big( \hat k_t /0.23 \big)^{3.3} & \qquad b_4 \text{ error}\,,
\end{array}\right. \label{eq:bispectrum_TE}
\eeq
where $\hat k_t \equiv \frac{1}{3}(k_1+k_2+k_3)/( h\,  {\rm Mpc}^{-1} )$.

\begin{figure}[t!]
\centering
\includegraphics[scale=1.]{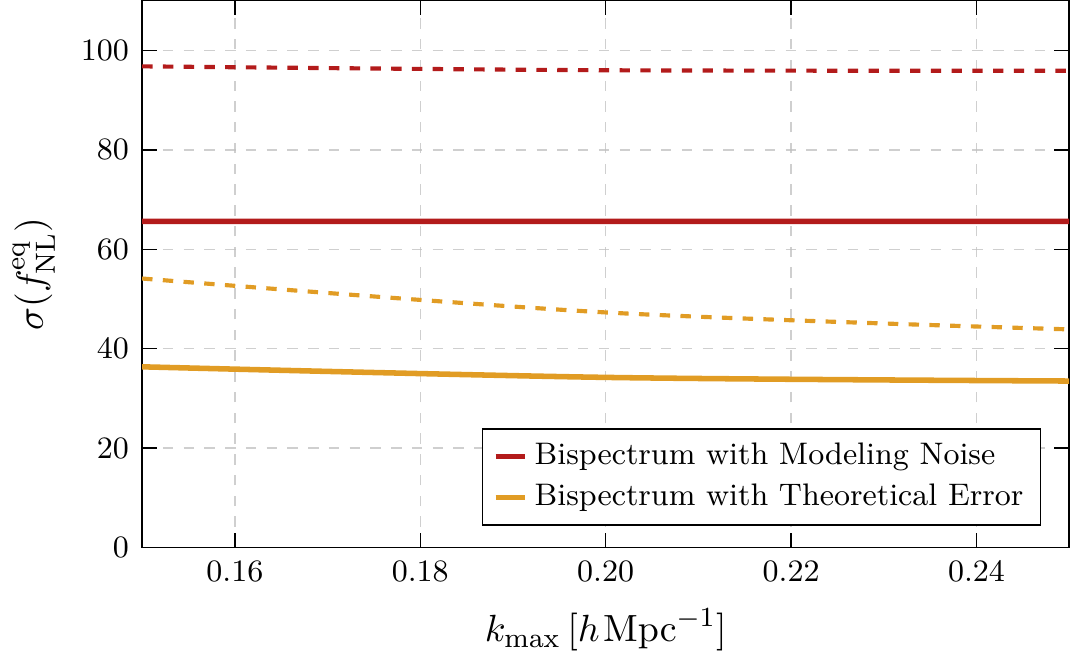}
\caption{ Bispectrum forecasts for $\sigma(\fnleq)$ as a function of $k_{\rm max}$ for a $300\, ({\rm Gpc}/h)^3$ survey with $10^{10}$\,objects at $z=0$.  These forecasts use $b_3$ theoretical error ({\it orange}) or modeling noise ({\it red}) and marginalize over $b_2, b_{{\cal G}_2}$ ({\it solid}) and $b_2, b_{{\cal G}_2}, b_{\partial^2}$ ({\it dashed}), with $R_*^{-1} = 0.3 \, h \, {\rm Mpc}^{-1}$.}
\label{fig:TEvMN}
\end{figure}

\vskip 10pt
\noindent
{\bf Modeling noise}---Looking at the expression for the bispectrum Fisher matrix~(\ref{eq:bispectrum_fisher}) suggests another form of theoretical error.
Concretely, the factor of $P(k) +N^2$ in~(\ref{eq:bispectrum_fisher}) represents the expected variance of each mode due to cosmic variance and stochastic bias $\epsilon(\x)$, whose two-point function we took to be $\langle \epsilon(\x) \epsilon(\x^{\hskip 1pt \prime}) \rangle = N^2 \delta_{\rm D}(\x-\x^{\hskip 1pt \prime})$. The field $\epsilon(\x)$ is meant to capture all contributions of $\dgalO(\x)$ that are not modeled by $\delta_g(\x)$.  Although we usually assume that the stochastic bias is dominated by shot noise, $N^2 \approx \bar n^{-1}$, the theoretical error is an additional source of disagreement between our model and the observed map $\dgalO(\x)$.  Making the same assumptions about the theoretical error that led to $V_{\rm eff}$ in~(\ref{eq:Veff_def}) 
 gives a $k$-dependent effective noise,\footnote{While the appearance of $C_e(k)$ is intuitive from~(\ref{eq:C_e}), the additional factor of $k^2 \Delta k V$ may require some explanation.  The theoretical error is local in $k$ and not $\k$ and is not assumed to be statistically isotropic.  A single $k$ in our expression actually represents $k^2 \Delta k V$ modes of momentum $\k$ with $|\k| \in( \k, \k+\Delta k)$ and the signal-to-noise of $P(k)$ at each $k$ is enhanced by the number of modes.  Since $P(k)/(P(k)+N^2)$ is the signal-to-noise per $\k$ and not $k$, the effective modeling noise per $\k$ must include this additional factor to compensate.}
\beq
P(k) + N^2 \to P(k) + N^2+k^2 \Delta k V C_e(k)\,,
\eeq
where $C_e(k)$, defined in~(\ref{eq:C_e}), is the power spectrum of the theoretical error (which is related to---but not exactly the same as---the theoretical error of the power spectrum).  We will refer to this additional term as {\it modeling noise} as it contributes contributes directly to the noise level of any correlation function.  

\vskip 4pt
Figure~\ref{fig:TEvMN} shows our results for bispectrum forecasts with both (bispectrum) theoretical errors and modeling noise. The qualitative behavior of both forecasts is similar, but the modeling noise makes a larger contribution to the effective noise and leads to a more significant degradation of $\sigma(\fnleq)$.
 The modeling noise results are also approximately independent of $k_{\rm max}$ because the higher effective noise means all modes with $k> 0.1 \, h \, {\rm Mpc}^{-1}$ are noise dominated.  The bispectrum theoretical error is a lower source of noise and produces smaller values of $\sigma(\fnleq)$.  The origin of the difference is that bispectrum theoretical error is making the stronger assumption that the source of uncertainty is only in the modeling of the bispectrum itself.  However, the covariance of the bispectrum measurement, which determines the noise, is sensitive to uncertainties in the map beyond the bispectrum and it is less clear that the dominant source of theoretical error in a realistic bispectrum analysis is necessarily given by~(\ref{eq:bispectrum_TE}).  The modeling error makes the more conservative assumption that the uncertainty is controlled by our uncertainty in the variance of the map.
  
\vskip 4pt
We note that it is essential for the results in this paper that neither the theoretical errors nor the modeling noise capture the local nature of the nonlinearity.  Rather than being local in real space, these errors are assumed to be local in Fourier space.  While this is a conservative assumption when nothing is known about the structure of the nonlinear effects, this assumption breaks the central physical principle that we believe enables improved constraints on $\fnleq$. Moreover, it is this assumption of locality in Fourier space that gives rise to an effective noise that grows with the volume of the survey.  As we saw in Figure~\ref{fig:Veff},  the constraints on $\fnleq$ therefore don't improve as fast as $\sigma(\fnleq) \propto V^{-1/2}$ with increasing volume.

\newpage
\subsection{Map-Level Forecasts} \label{sec:map_forecasts}

We now want to compare these bispectrum forecasts to a map-level analysis.
Moreover, we will extend the forecasts to include the impact of $b_{n>2}$ on $\fnleq$.
The relevant Fisher matrix elements were given in (\ref{equ:FisherWith_Obs}):
\beq
\begin{aligned}
F_{n,m} &\ =\ V  \int   \frac{\d^3 k}{(2\pi)^3}  \frac{  \langle [ (\dgalO)^n](\k) [(\dgalO)^m](-\k) \rangle' + n \delta_{n,m} n! \sigma^{2n}}{P(k) + N^2} \,,\\[8pt]
F_{n, \rm eq} &\ =\  V \sum_m \int  \frac{\d^3 k}{(2\pi)^3} \, m \bar b_m \frac{\langle [(\dgalO)^n](\k)[(\dgalO)^{m-1}\delta_{\rm NG}[\dgalO] ](-\k) \rangle'}{P(k)+N^2}   \\
& \quad\ \ \ + \frac{2}{3}\delta_{n,2} \int \frac{\d^3 p}{(2\pi)^3} \frac{B_{\rm eq}(\p,\k-\p\hskip 1pt)}{(P(|\k-\p\hskip 1pt |)+N^2) (P(k)+N^2)} \ , \\[8pt]  
F_{\rm eq, eq} &\ =\ V \int  \frac{\d^3 k}{(2\pi)^3}  \frac{\langle \delta_{\rm NG}[\dgalO](\k) \delta_{\rm NG}[\dgalO](-\k)\rangle' }{P(k)+N^2}\\
&\quad\ \ \ +  \frac{V}{9} \int \frac{\d^3 p}{(2\pi)^3} \frac{B_{\rm eq}(\p,\k-\p\hskip 1pt ) B_{\rm eq}(-\p, \k)}{(P(|\k-\p\hskip 1pt |)+N^2) (P(p)+N^2) (P(k)+N^2)} \, ,
\end{aligned}
\eeq
where, by an abuse of notation, we defined $\langle \ldots  \rangle'$ as the statistical average of possible observed maps $\dgalO(\x)$.  In practice, this means that we are taking $\bar \delta(\x) \to \delta(\x)$ in $\dgalO(\x)$ and use the erodic theorem to calculate the Fisher matrix. We will evaluate these expressions to leading order in $\delta(\x)$, so that in most cases they reproduce~(\ref{equ:FisherWith}).  However, some off-diagonal terms that vanish in~(\ref{equ:FisherWith}) get a nonzero, but significantly suppressed contribution, in our forecasts.  
To provide a direct comparison with the theoretical error estimates in~(\ref{eq:bispectrum_TE}), we use the same power-law scalings to determine these contributions: 
\begin{align} \label{eq:b3b3}
\langle [(\dgalO)^3](\k) [(\dgalO)^3](-\k) \rangle' &\approx b_1^6 \big(\hat k/0.31\big)^{1.8} \langle [\delta^2(\k)] [\delta^2](\k') \rangle'  \,, \\[4pt]
\langle [(\dgalO)^3](\k) [(\dgalO)^2](-\k) \rangle' &\approx b_1^4 b_2 \big(\hat k/0.31 \big)^{1.8}  \langle [\delta^2(\k)] [\delta^2](\k') \rangle'  \label{eq:b3b2} \, ,
\end{align}
where we have included factors of $b_1$ and $b_2$ for clarity. We will apply this scaling behavior for the mixing of $b_3$ with $\fnleq$, $b_2$, $b_{{\cal G}_2}$. The scaling behavior matches~(\ref{eq:scaling_fisher}) except that $\langle [\delta^2(\k)] [\delta^2](\k') \rangle'$ is evaluated in $\Lambda$CDM and not the scaling universe.  For the contributions from $b_4$ ($b_5$), we use the same expressions except that the spectral index is changed from $1.8 \to 3.3$ ($1.8 \to 5.1$) and the reference scale from $0.31 h \, {\rm Mpc}^{-1} \to 0.23 \, h \, {\rm Mpc}^{-1}$.  The contributions to the Fisher matrix for $b_3$, $b_4$, $b_5$ are therefore just a $k$-dependent rescaling of the Fisher matrix element for $b_2$  calculated using the bispectrum (\ref{equ:B2}).

\vskip 4pt
Our approach to modeling the higher-order contributions to the map is approximate, but fairly conservative.  As we saw in Figure~\ref{fig:cosines}, the quadratic bias parameter $b_2$ is highly degenerate with $\fnleq$ and therefore our approximation of the contributions from $b_3$, $b_4$, $b_5$ only account for the suppression from $k < \knl$ that we would anticipate at higher orders, but we are otherwise not making any special assumptions that would artificially reduce the degeneracy.  This approach is, of course, insufficient for analyzing maps of the real universe and the inaccuracy in the model would bias the measurement of $\fnleq$.  However, at the level of a Fisher analysis, we are always assuming that the data is generated from the same model as our theory (plus stochastic bias) and therefore the issue of bias is irrelevant.  For the purpose of understanding the degeneracy between late-time nonlinearities and primordial non-Gaussianity, there is no reason to expect any meaningful changes to the results in this section were we to replace~(\ref{eq:b3b3}) and~(\ref{eq:b3b2}) with the exact one-loop expressions.

\vskip 10pt
\noindent
Using the above forecasting methodology, we can now compare the map-level forecasts and the more traditional bispectrum forecasts.  
To isolate the role of nonlinearities in the forecasts,
we will work at a fixed redshift $z=0$ with varying survey volumes and $k_{\rm max}$. 

\begin{figure}[t!]
\centering
\includegraphics[scale=1.]{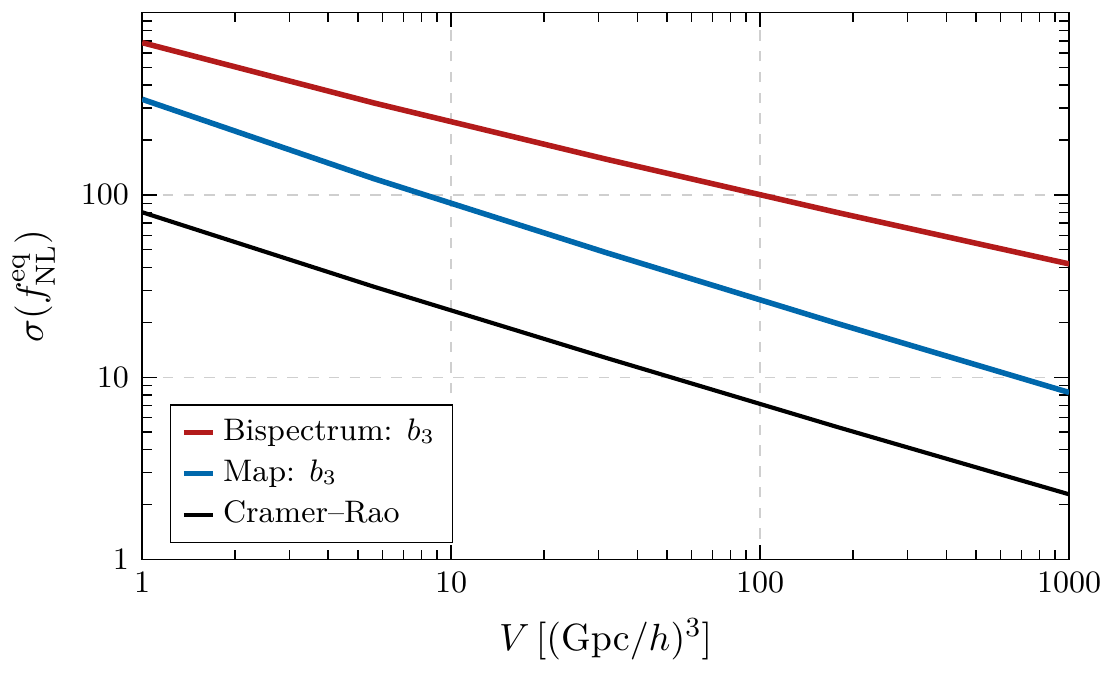}
\caption{Forecasts for $\sigma(\fnleq)$ as a function of the survey volume, for a $z=0$ survey with $10^{10}$\,objects and $k_{\rm max} = 0.2 \, h \, {\rm Mpc}^{-1}$. We show the result of the map-level forecasts marginalized over $b_3$ ({\it blue}) compared to bispectrum forecasts with $b_3$ modeling noise ({\it red}) and the Cramer--Rao bound ({\it black}).}
\label{fig:Vol}
\end{figure}

\subsubsection*{Results}

Figure \ref{fig:Vol} illustrates the potential of map-based forecasts to improve constraints on $\fnleq$ as a function of the survey volume  for fixed $k_{\rm max} = 0.2 \, h \, {\rm Mpc}^{-1}$.  Within our simplified forecasting framework, this gives a sense how achievable improvements in $\sigma(\fnleq)$ can be with future surveys.  We have taken the number of objects to be fixed at $10^{10}$ to avoid the impact of shot noise except for very large volumes.  While this is unrealistic for planned LSS surveys, the effective shot noise of 21\hskip 1pt cm intensity mapping surveys like PUMA~\cite{CosmicVisions21cm:2018rfq} can be consistent with such a large effective number of objects.   The map-level forecasts for $\fnleq$ are calculated after marginalizing over $b_2$, $b_{{\cal G}_2}$, $b_3$ and are compared to the bispectrum forecasts with marginalization over  $b_2$, $b_{{\cal G}_2}$ and with $b_3$ modeling noise.  The Cramer--Rao line shows the optimal constraint on $\fnleq$ in these surveys, with no marginalization over additional parameters. At $V = 300 \, ({\rm Gpc}/h)^3$, we find $\sigma(\fnleq) = 66, 15,$ and $4.0$ for the bispectrum, map-level and Cramer--Rao forecasts, respectively.  We caution again that these numbers are for an artificial $z=0$ survey and would improve when higher redshift information is included. We also see that the map-level constraints improve, relative to the bispectrum constraints, with increasing volume, reflecting the difference in the effective volumes illustrated in Figure~\ref{fig:Veff}.  In this sense, not only are the map-level forecasts more optimistic in terms of raw sensitivity, they also show that we recover the expected improvements from increasing the volume, in contrast to the bispectrum forecasts with modeling noise.

\begin{figure}[t!]
\centering
\includegraphics[scale=1.1]{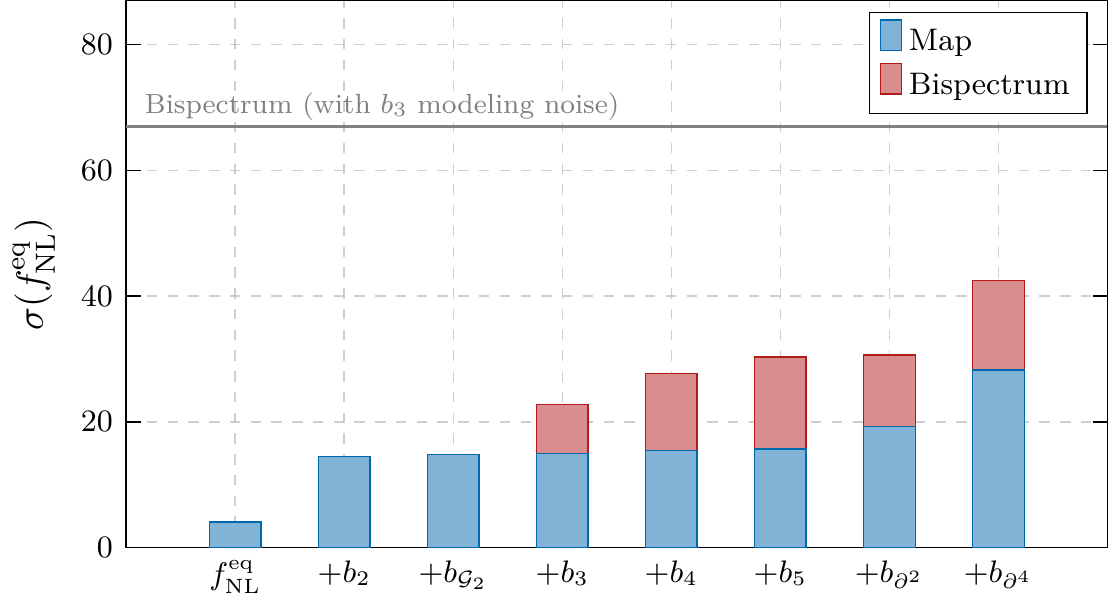}
\caption{Summary of forecasts for a 300 $({\rm Gpc}/h)^3$ survey with $10^{10}$\,objects and $k_{\rm max} = 0.2\, h \, {\rm Mpc}^{-1}$ at $z=0$.  As before, we work at $z=0$ to highlight the improved mitigation of nonlinear effects, at the cost of increasing the values of $\sigma(\fnleq)$ compared to a realistic, higher-$z$ survey.  The leftmost bar, $\fnleq$, is for the case without nonlinear biasing, so that $\sigma(\fnleq)=\sigma(\fnleq)_{\rm min}$.  Moving from left to right includes marginalizing over each additional parameter.  The blue bars are for map-level forecasts while the red bars are for bispectrum forecasts with no theoretical errors.  The gray line is the bispectrum forecast marginalized over $\{b_2, b_{{\cal G}_2}\}$ with $b_3$ modeling noise (the same as in Fig.\,\ref{fig:TEvMN}). We see that increasing the number of higher-order local bias parameters ($b_n>2$) has no effect in the map-level forecasts, but increases  $\sigma(\fnleq)$ in the bispectrum forecasts.  In contrast, both types of forecasts degrade when higher-derivative terms are added, consistent with the breakdown of locality.} 
\label{fig:b5}
\end{figure}

\vskip 4pt
In Figure~\ref{fig:b5}, we compare the map-based forecasts to the bispectrum forecasts {\it without modeling noise}.\footnote{In the case without modeling noise, we assume that the bispectrum can be modeled to any needed accuracy if we include enough bias parameters.  Our inability to model the physics of structure formation perfectly, despite the knowledge that it is local, is then captured by marginalizing over these bias parameters (with flat priors). }   The figure illustrates the effect of including marginalization over additional bias parameters.  
The leftmost bar, labeled $\fnleq$, is for the case without any nonlinear biasing and therefore describes the optimal constraint $\sigma(\fnleq)=\sigma(\fnleq)_{\rm min}$. 
The next two bars, labeled $+b_2$ and $+b_{{\cal G}_2}$, show how this constraint degrades when marginalizing over the quadratic bias parameters.  As we discussed previously, the bispectrum and map-level constraints are equivalent for quadratic biasing, so this degradation is the same for both methods.
The power of the map-level analysis becomes apparent when we include additional higher-order bias parameters (here $b_3$, $b_4$ and $b_5$).  We see that the map-level constraints remain largely unaffected, while the bispectrum constraints degrade significantly. Note that the bispectrum constraints are optimistic in the sense that the assume {\it no} theoretical errors and a relatively large $k_{\rm max} = 0.2\,h{\rm Mpc}^{-1}$. For comparison, we also show the bispectrum constraint with $b_3$ modeling noise.

\vskip 4pt
In the bispectrum analysis, marginalizing over the higher-order bias parameters
 $b_3$, $b_4$ and $b_5$ is equivalent to marginalizing over the one-, two-, and three-loop contributions.
Since the shape of these contributions is degenerate with equilateral non-Gaussianity, this degrades the constraint on~$\fnleq$.
 In contrast, in the map-level analysis, the constraints are insensitive to marginalization over these additional bias parameters. Effectively, the map implicitly contains information from higher-point correlations---up to the six-point function in this example---which breaks the degeneracies between $\fnleq$ and $b_{n>2}$.\footnote{ When we calculate the individual correlation functions in Fourier space, the constraints imposed by locality are not hardwired into the calculation and are even broken by the modeling errors (which are local in Fourier space rather than position space).  Enforcing locality at the map level imposes relations between correlators that ensures that these higher-order nonlinear terms do not mix with the primordial signal.}  As a map-based statement, it is simply that $b_{n>2}$ produces a localized change to the map that is easily distinguished from the effect of $\fnleq$.
 
 \begin{figure}[t!]
\centering
\includegraphics[scale=1.]{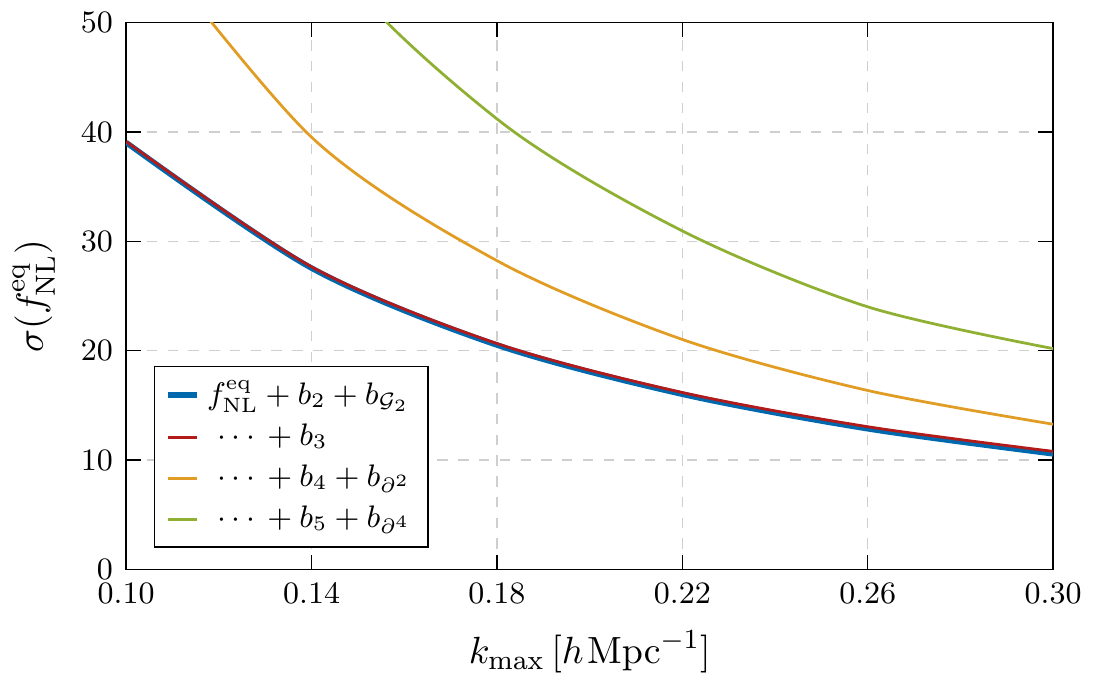}
\caption{Forecasts for the map-level constraints as a function of $k_{\rm max}$ for the same survey as in Fig.\,\ref{fig:b5}. Note that the forecasts improve significantly with increasing $k_{\rm max}$, in contrast to the bispectrum forecasts with $b_3$ theoretical error or modeling noise in Fig.\,\ref{fig:TEvMN}.}
\label{fig:kmax}
\end{figure}

\vskip 4pt
What protects $\fnleq$ from significant degeneracies with nonlinear biasing is the local nature of the late-time evolution.   As a result, the central challenge with measuring primordial non-Gaussianity at the map level is control over nonlocal effects, which are captured by higher-derivative contributions in the bias expansion, like $R_*^2 \partial^2 \delta$ and $R_*^4 \partial^4 \delta$.  We see, in Figure~\ref{fig:b5}, that both the map-level and bispectrum constraints indeed weaken when including marginalization over these higher-derivative  terms. 
The dependence on $k_{\rm max}$ of these results is illustrated in Figure~\ref{fig:kmax}.
Unlike the bispectrum forecasts in Fig.\,\ref{fig:TEvMN}, the map-level forecasts improve significantly with increasing  $k_{\rm max}$.
We see from these results that the most important aspect of the nonlinear modeling is to capture the range of possible nonlocal evolution.  When we add higher-order nonlocal terms with arbitrary coefficients, we are allowing for arbitrary nonlocal evolution (this is just the Taylor expansion of an unknown nonlocal function).  Locality on the scale $R_*$ should be more hardwired into the analysis as it is a fundamental aspect of the physical evolution and the origin of the protection of the primordial non-Gaussianity.

\subsubsection*{Summary}

A key take-away from these results is that for local nonlinear effects, the degeneracy between $b_2$ and $\fnleq$ is the main source of the increase in $\sigma(\fnleq)$ relative to a $\fnleq$-only forecast.  In the absence of nonlocal higher-derivative operators, a forecast with only quadratic biasing captures all of the map-level information. We also showed that in this case the results are equivalent to a bispectrum-only forecast with $b_2$ and $\fnleq$ included (and {\it no} theoretical errors).  We can therefore translate existing bispectrum forecasts on $\fnleq$ from planned surveys into map-level forecasts provided these forecasts marginalized over $b_2$ (and neglected theoretical error).  Such forecasts for near-term optical surveys like Euclid give $\sigma(\fnleq) = 7.5$ for $k_{\rm max} = 0.15 \, h \, {\rm Mpc}^{-1} / D(z)$~\cite{Amendola:2016saw,Giannantonio:2011ya} and  $\sigma(\fnleq) = 16$ for $k_{\rm max} = 0.1 \, h \, {\rm Mpc}^{-1} / D(z)$~\cite{Karagiannis:2018jdt}.  Since the map-level forecasts are more robust to nonlinearities, it seems feasible to use $k_{\rm max} = 0.2 \, h \, {\rm Mpc}^{-1} / D(z)$ in these surveys, which would significantly improve the constraints.\footnote{The PUMA survey~\cite{CosmicVisions21cm:2018rfq} suggests that for optimistic foreground removal, $\sigma(\fnleq) = 4.5$ would be achievable given their large survey volume and low effective shot noise even {\it with} theoretical error and using a conservative $k_{\rm max} = 0.1 \, h \, {\rm Mpc}^{-1} / D(z)$.  It would in interesting to see if $\sigma(\fnleq) \lesssim 1$ is achievable when relaxing these assumptions.} 
\vskip 4pt
These forecasts suggest room for significant improvements over the CMB in the measurements of $\fnleq$ with future LSS surveys.  A conventional point of view has been that even these conservative forecasts are likely under-estimating the impact of nonlinearities.  However, given the rise of map-based forward modeling and the power of locality, the potential of these surveys should be revisited from a less pessimistic point of view.

\section{Outlook and Conclusions}\label{sec:conclusions}

Locality is a key principle that underlies our understanding of the physical world.
Modern physics is formulated in terms of relativistic field theories that encode local interactions and determine their causal influences between separated points in spacetime.  However, at the level of observables, the constraints imposed by locality and causality are not always fully transparent.
For example, in cosmological perturbation theory we typically work in Fourier space where locality and causality aren't manifest, but instead are encoded in the analytic structure of correlation functions.  In addition, the local nature of interactions implies specific relations between individual correlation functions.  So far, the benefit of perturbative calculations being simpler in Fourier space has outweighed the cost of losing direct manifestations of locality constraints. However, in future large-scale structure studies, we are sensitive to the breakdown of perturbation theory, where simulation-based methods are unavoidable, so that it may be worth revisiting our reliance on these perturbative methods.

\vskip 4pt
In this paper, we have explored what can be gained by a position space, map-level treatment of large-scale structure correlations. A critical feature of inflation is that it produces correlations between spatially separated points that appear to be {\it nonlocal} for a late-time observer---this is an incarnation of the horizon problem of the hot Big Bang cosmology. Meanwhile,  due to the slow speed of propagation of matter after recombination, the nonlinear evolution of structure in the late universe is necessarily local. 
Because of locality, late-time nonlinearities at a fundamental level therefore {\it cannot} mimic the inflationary non-Gaussianities.  However, when these signatures are written only in terms of the bispectrum of the local density, the vastly different character of the correlations is lost. We showed that by working instead with the complete position space maps, this essential distinction becomes manifest again and opens the door to improved measurements of non-Gaussianity in upcoming large-scale structure surveys.

\vskip 4pt
The map-level analysis introduces information beyond the bispectrum into our constraints on non-Gaussianity and breaks the degeneracy between higher-order nonlinearities and the primordial signals.  This effectively reduces the  potential sources of bias or degeneracy from an infinite list of higher-order terms, down to the leading nonlinear terms.  In the case of galaxy clustering, the main source of degeneracy are the quadratic bias coefficients $b_2$ and $b_{{\cal G}_2}$.  In principle, even this degeneracy could be reduced using information from the power spectrum and cross correlations (perhaps effectively playing the role of cosmic variance cancelation~\cite{Seljak:2008xr}). 

\vskip 4pt
While the results of this paper suggest an optimistic outlook (particularly in comparison with previous expectations), a concrete demonstration that these improvements are achievable in realistic surveys remain an important open problem.  First of all, our forecasts did not include complications associated with redshift space distortions or with bulk flows and reconstruction.  While a number of prior works suggest that these additional effects can be mitigated, if not removed entirely, they do involve large velocities that could potentially interfere with the protection from locality.  In addition, the map-level analysis will rely on fast simulation-based techniques~\cite{Taylor:2019mgj,Dai:2020ekz,Modi:2020dyb,Modi:2021acq,Makinen:2021nly,Hassan:2021ymv,Villaescusa-Navarro:2021pkb,Villaescusa-Navarro:2021cni} that are still under development and possible limitations of this approach for primordial non-Gaussianity may not be fully understood yet.  None of these issues seems to present a fundamental obstacle to measuring non-Gaussianity at the map level, but this needs to be demonstrated.

\vskip 4pt
Although the focus of this work was on the specific case of equilateral non-Gaussianity, the benefits of locality should apply much more broadly.  In fact, in a sense the case of equilateral non-Gaussianity is a ``worst case scenario" because the bispectrum of the late-time nonlinearities is highly degenerate with the equilateral bispectrum shape. We therefore expect locality considerations to be even more powerful for other inflationary signals. Indeed,  for some of the signatures~\cite{Chen:2009zp,Baumann:2011nk,Assassi:2012zq,Noumi:2012vr,Arkani-Hamed:2015bza,Lee:2016vti} associated with ``cosmological collider physics" the nonlocal physics that protects the signal is much more pronounced and already allows meaningful constraints from the power spectrum~\cite{Gleyzes:2016tdh}. This opens the door to more general map-based non-Gaussianity searches than have previously been considered with LSS data.

\paragraph{Acknowledgements}

We are grateful to Matteo Biagetti, Giovanni Cabass, Alex Cole, Mikhail Ivanov, Mehrdad Mirbabayi, Rafael Porto, Alec Ridgway, Eva Silverstein, Marko Simonovi\'c, An\v ze Slosar, Ben Wallisch and Matias Zaldarriaga for helpful discussions. DG is supported by the US~Department of Energy under Grants~\mbox{DE-SC0019035} and~\mbox{DE-SC0009919}.
DB receives funding from a VIDI grant of the Netherlands Organisation for Scientific Research~(NWO) and is part of the Delta-ITP consortium.  
DB is also supported by a Yushan Professorship at National Taiwan University funded by the Ministry of Science and Technology (Taiwan). 

\newpage
\appendix

\section{From Likelihoods to Initial Conditions}\label{app:likelihoods}

In this appendix, we will fill in some details of the computations presented in Section~\ref{sec:map}.
Our starting point is the likelihood function
\begin{align}
{\cal L} &= \exp \left( - \int \d^3 x\hskip 1pt \d^3 x'\, \delta(\x) \hskip 1pt C_2(\x-\x^{\hskip 1pt \prime})  \hskip 1pt \delta(\x^{\hskip 1pt \prime}) - \frac{1}{2N^2} \int \d^3 x \left(\dgal(\x) - \dgalO(\x) \right)^2 \right) ,
\end{align}
where $\dgalO$ is the observed galaxy density contrast and $\delta$ the linearly evolved matter density, which depends on the unknown Gaussian initial conditions.  
Recall that the observed galaxy map is $\dgalO(\x) = \delta_g(\x) +\epsilon(\x)$, where $\delta_g(\x)$ is the model for the galaxy map  and $\epsilon(\x)$ is the stochastic bias. The model depends on the bias parameters $b_n$ and possibly the amplitude $ \fnleq$ of the non-Gaussian initial conditions.
Our goal is to find  the parameters $\{ b_n, \fnleq\}$ and the linear map $\delta(\x)$ that maximize the likelihood.

\subsection{Maximum Likelihood Map}

 As described in Section~\ref{sec:Section3-2}, the maximum likelihood map can be found
 iteratively, by first
 finding the maximum likelihood values of the model parameters $\{b_n,\fnleq\}$, for some initial fixed guess at the Gaussian map,~$\bar \delta_{\rm guess}$, and then determining the maximum likelihood solution of the map, $\dbarO(\x)$, for fixed model parameters.

\vskip 4pt
We hence start with some fixed $\bar \delta_{\rm guess}(\x)$, although we anticipate that $\bar \delta_{\rm guess}(\x) \to \dbarO(\x) \approx  \bar \delta(\x)$ will result from our iterative procedure.
Taking a derivative of $-\log {\cal L}$ with respect to $b_m$ then gives  
\begin{align}
 -\frac{\d}{\d b_m} \log {\cal L} = \frac{1}{N^2} \int \d^3 x \, (\bar \delta_{\rm guess})^m(\x) \left(\dgal(\x) - \dgalO(\x) \right)=0\,,
\end{align}
where 
\begin{align}
\dgal(\x)  &\approx \sum_n b_n (\bar \delta_{\rm guess})^n(\x) + b_1 \fnleq  \bar \delta^{\rm guess}_{\rm NG}(\x)    \,,\\
 \dgalO(\x) &\approx \sum_n \bar b_n \bar \delta^n(\x)  + \bar b_1 \bar f_{\rm NL}^{\rm eq} \bar \delta_{\rm NG}(\x)  + \epsilon(\x) \,,
 \end{align}
 with  $\{\bar b_n, \bar f_{\rm NL}^{\rm eq} \}$ the fiducial values and
 \beq
\fnleq \bar \delta^{\rm guess}_{\rm NG}(\k) 
= \int \frac{\d^3 p \hskip 1pt \d^3 q}{(2\pi)^3}\frac{B_{\rm eq}(\vec p,\vec q\hskip 1pt)}{P(p)P(q) } \,\bar \delta_{\rm guess}(\p\hskip 1pt) \bar \delta_{\rm guess}(\q\hskip 1pt)\, (2\pi)^3 \delta_{\rm D}(\p+\q-\k)\, .\label{eq:app_dNG}
\eeq
{\it If} we assume that $\bar \delta_{\rm guess}(\x) \approx \bar \delta(\x)$, then we can use the ergodic theorem, so that  
\beq
\begin{aligned}
\int \d^3 x\, \bar \delta^m(\x)\bar \delta^n (\x) &\to V \langle \delta^m(\x) \delta^n (\x)  \rangle \, ,\\
\int \d^3 x\, \bar \delta^m(\x) \epsilon (\x) &\to V \langle \delta^m(\x) \epsilon (\x)  \rangle =0\,.
\end{aligned}
\eeq
It is then easy to see that the maximum likelihood points are $b_n = \bar b_n$ and $\fnleq = \bar f_{\rm NL}^{\rm eq}$. This shows that {\it if} the iterative procedure approaches the correct initial conditions, then the measured model parameters will agree with the fiducial parameters.  

\vskip 4pt
Our primary goal is to see that we can determine the initial conditions accurately.  We will now assume that the model parameters are fixed near their fiducial values (we can further iterate after making an estimate of the map of the initial conditions). To determine the maximum likelihood solution for the Gaussian map,  we take a derivative of $-\log {\cal L}$  with respect to $\delta(\x$), so that
\beq
\begin{aligned}
0 = -\frac{\d}{\d \delta(\x)} \log {\cal L} &= \int \d^3 x' \,C_2(\x-\x^{\hskip 1pt \prime}) \delta(\x^{\hskip 1pt \prime}) + \frac{1}{N^2} \left(\sum_n n b_n \delta^{n-1}(\x) \right) \left(\dgal(\x) - \dgalO(\x) \right) 
\\
&\quad + \frac{1}{N^2}  \int \d^3 x' \,\frac{\d \hskip 1pt \delta_{\rm NG}(\x^{\hskip 1pt \prime})}{\d \delta(\x) } \left(\dgal(\x^{\hskip 1pt \prime}) - \dgalO(\x^{\hskip 1pt \prime}) \right) .
\end{aligned}
\eeq
Note that the naive solution $\dgal(\x) = \dgalO(\x)$ does {\it not} maximize the likelihood.  The optimal solution should minimize the combination of cosmic variance and shot noise, and the former is not minimized by $\dgal(\x) = \dgalO(\x)$.  
As we explained in Section~\ref{sec:Section3-2},
we can find a maximum likelihood solution for the ``observed" Gaussian initial conditions, $\dbarO(\x)$, as an inversion of the observed map 
\begin{align}
\dbarO(\x) &= \sum_{n} \tilde b_n \star [(\dgalO)^n] (\x) - \fnleq  \bar \delta_{\rm NG}[\dgalO](\x)\, , \label{equ:ObsFilter}
\end{align}
where $f \star g(\x)$ denotes a convolution and $\bar \delta_{\rm NG}$ is a function which removes the primordial non-Gaussianity from the map.  When the filter functions $\tilde b_n(\x)$ are proportional to a delta function, then (\ref{equ:ObsFilter}) reduces (\ref{equ:delta}), i.e.~it becomes the inversion of the map in the absence of noise.  For realistic data, we need to include the filters to down-weight the impact of noisy modes. Without the filter functions, the variance of the reconstructed $\dbarO$ would follow $\dgalO$ even in the noisy regime where $N^2 \gg P(k)$.  Since the maximum likelihood solution, $\dbarO(\x)$, is our best determination of the true Gaussian initial condition, it must also appear in the model for the galaxy density that we compare to observations,
\begin{align}
\delta_g(\x) &= \sum b_n  \left[\left(\dbarO(\x)+ \fnleq \delta_{\rm NG}[\dbarO](\x) \right)^n \right] ,
\end{align}
where $\delta_{\rm NG}$ is the same as~(\ref{eq:app_dNG}) evaluated with the maximum likelihood solution $\dbarO(\x)$ in place of $\bar \delta(\x)$.  

\vskip 4pt
In Fourier space, the convolutions in (\ref{equ:ObsFilter}) simply become multiplications by the Fourier transform of the filter functions, $\tilde b_n(\k)$.  At linear order in $\dgalO$, the maximum likelihood solution determines $\tilde b_1(\k)$ alone:
\beq
\dgalO(\k) \left( \frac{\tilde b_1(\k)}{P(k)} +  \frac{b_1}{N^2}\left(b_1 \,  \tilde b_1(\k) - 1\right) \right)= 0 \quad \to \quad  \tilde b_1(\k) = \frac{b_1 P(k)}{b_1^2 P(k)+N^2} \, .
\eeq
This result makes intuitive sense: At linear order, the best-fit Gaussian field is just a noise-filtered version of the observed map.  

\vskip 4pt
At second order in $\dgalO$, we determine both $\tilde b_2(\k)$ and the leading contribution to~$\bar \delta_{\rm NG}$.  We find $\tilde b_2(\k)$ by dropping terms proportional to $\fnleq$:
\beq
\tilde b_2(\k)\frac{[(\dgalO)^2](\k)}{P(k)} + \frac{1}{N^2} \left(b_1 \tilde b_2(\k) [(\dgalO)^2](\k) + b_2 [(\tilde b_1\star \dgalO)^2](\k) \right) = 0\,,
\eeq
which gives
\begin{align}\label{eq:delta2}
\tilde b_2(\k) &= - b_2 \frac{b_1 P(k)}{b_1^2 P(k)+N^2} \frac{[(\tilde b_1\star \dgalO)^2](\k)}{[(\dgalO)^2](\k)} \ \xrightarrow{\ P(k) \gg N^2\ } \ - \frac{b_2}{b_1^2} \frac{b_1 P(k)}{b_1^2 P(k)+N^2}  \, .
\end{align}
In the limit of high signal-to-noise, this simply removes the ${\cal O}((\dgalO)^2)$  error in $\dbarO$ introduced  by our ${\cal O}(\dgalO)$ solution for $\tilde b_1(\k)$. 
For $\fnleq \neq 0$, we determine $\bar \delta_{\rm NG}[\dgalO]$ at ${\cal O}((\dgalO)^2)$ and ${\cal O}(\fnleq)$ by solving  
\beq
\fnleq \left( - \frac{ \bar \delta_{\rm NG}[\dgalO](\k)}{P(k)} + \frac{b_1^2}{N^2}\left(-  \bar \delta_{\rm NG}[\dgalO](\k) + \delta_{\rm NG}(\tilde b_1 \star \dgalO)(\k)\right) \right) = 0\,,
\eeq
so that 
\beq
\bar \delta_{\rm NG}[\dgalO](\k)= \frac{b_1^2 P(k)}{b_1^2 P(k)+N^2} \hskip 2pt \delta_{\rm NG}(\tilde b_1 \star \dgalO)(\k) \,.
\eeq
At this order, we are removing the noise-weighted non-Gaussian term evaluated using the first-order solution for $\dbarO$.

\subsection{Higher-Order Filtering}
\label{app:beyond_linear}

Having developed some intuition for the form of the filter functions, we now determine the higher-order contributions needed to calculate the Fisher matrix.  To avoid clutter, we will set $b_1 \equiv 1$.  

\vskip 4pt
To determine $\tilde b_3(\k)$, we have to work up to cubic order in $\dgalO$:
\beq\label{eq:tb3}
\begin{aligned}
\tilde b_3(\k)\frac{[(\dgalO)^3]}{P(k)} &+ \frac{1}{N^2} \left(\tilde b_3(\k) [(\dgalO)^3]+ b_3 [(\tilde b_1\star \dgalO)^3]+ 2 b_2  \dgalO \star \tilde b_2  \star [(\dgalO)^2] \right)\\
& + 2 b_2 \,\tilde b_1 \star \dgalO \star b_2 \frac{1}{P(k)+N^2} [(\dgalO)^2]= 0\, .
\end{aligned}
\eeq
Note that because the products are local in position space, we have used convolutions in momentum space and dropped the momentum argument.  Our goal is to match the coefficients of the operators, however, we notice that not all terms in~(\ref{eq:tb3}) are written in terms of renormalized operators.  Before solving for $\tilde b_3$, we first need to remove the self-contractions, namely 
\beq
\dgalO \star [(\dgalO)^2] = [(\dgalO)^3] + 2 \dgalO \sigma^2\,,
\eeq
where $\sigma^2  = \int \d^3 k\, P(k) /(2\pi)^3$ is the contraction of two $\dgalO$. 
We can absorb $\dgalO \sigma^2$ into a shift of the first-order filter function
\beq
\tilde b_1(\k) \to \frac{P(k)}{P(k)+N^2}+  4 \left(\frac{P(k)}{P(k)+N^2} \right)^2  \sigma^2 b_2^2\,,
\eeq
where we recall that we have set $b_1 =1$.  We then have 
\beq
\tilde b_3(\k)= - \frac{P(k)}{P(k)+N^2}\,b_3  +  2 \left(\frac{P(k)}{P(k)+N^2} \right)^2 b_2^2\, .
\eeq
In practice, only the shift of $\tilde b_1(\k)$ will directly impact the Fisher matrix at leading order, which is one of the benefits of working with renormalized operators. 

\vskip 4pt
Repeating the same procedure with the $(\fnleq)^2$ and $b_2 \fnleq$ terms, we again find that we have to shift $\tilde b_1(k)$ in order to write the expansion in terms of renormalized operators.  In the limit of high signal-to-noise, $P(k) \gg N^2$, we get
\beq
\begin{aligned}
\tilde b_1(\k) = 1 + 4 b_2^2 \sigma^2 &+ \frac{2}{3} \fnleq b_2\int \frac{\d^3 p}{(2\pi)^3} \frac{B_{\rm eq}(\p,\k-\p\hskip 1pt)}{P(|\k-\p\hskip 1pt |)} \\
&+ \frac{1}{9} (\fnleq)^2 \int \frac{\d^3 p}{(2\pi)^3} \frac{B_{\rm eq}(\p,\k-\p\hskip 1pt) B_{\rm eq}(-\p, \k)}{P(|\k-\p\hskip 1pt |) P(p) P(k)} \,. 
\end{aligned}
\eeq 
Although there are also ${\cal O}((\dgalO)^3)$ terms that enter $\bar \delta_{\rm NG}$, just like the $b_2^2$ contributions to $\tilde b_3(\k)$, these will not contribute to the Fisher matrices and so we did not write them out explicitly.

\subsection{Details of the Fisher Matrix}

Given the maximum likelihood maps, we can now define the Fisher matrix by 
\beq
F_{i j} = - \frac{\partial^2}{\partial \theta_i \partial \theta_j} \log {\cal L} \Big|_{b_n =\bar b_n, \fnleq=0}\,, 
\eeq
where 
\beq
\begin{aligned}
{\cal L} = \exp \bigg( & - \int \d^3 x\hskip 1pt \d^3 x'\, \dbarO(\x) \hskip 1pt C_2(\x-\x^{\hskip 1pt \prime})  \hskip 1pt \dbarO(\x^{\hskip 1pt \prime}) \\
& - \frac{1}{2N^2} \int \d^3 x \left(\dgal(\x) - \dgalO(\x) \right)^2 - {\rm Tr} \log \frac{\partial \dbarO}{\partial \dgalO}\bigg)\,  .
\end{aligned}
\eeq
Here, $\dbarO(\x)$ and $\delta_g(\x)$ are functions of $\{b_n, \fnleq\}$, with $\dgalO(\x)$ held fixed, and we have included the Jacobian from the change of variables from $\dbarO$ to $\dgalO$ following~\cite{Creminelli:2006gc}.   

\vskip 4pt
To illustrate the impact of the nonlinear terms in the Fisher matrix, let us first calculate $F_{2,2}$.
Evaluating this expression at high signal-to-noise, we find  
\begin{align}
F_{2,2} &= - \frac{\partial^2}{\partial b_2^2} \log {\cal L} \Big|_{b_n =\bar b_n, \fnleq=0} \nonumber \\[12pt]
&=
 \int \frac{\d^3 k}{(2\pi)^3}\bigg[ \frac{ P^2(k)  [(\dgalO)^2](\k) \, [(\dgalO)^2](-\k)}{P(k) (P(k)+N^2)^2} \nonumber+ \frac{1}{N^2} \left| \frac{P(k)}{P(k)+N^2} [(\dgalO)^2](\k) -  [(\tilde b_1 \star \dgalO)^2](\k) \right|^2 \nonumber \\
&\qquad \qquad \qquad + 8 \left(\frac{P(k)}{P(k)+N^2} \right)^2 \sigma^2 \frac{\dgalO(\k)\dgalO(-\k)}{P(k)+N^2} - 4 \sigma^2 \left(\frac{P(k)}{P(k)+N^2} \right)^2  \bigg] \nonumber \\[12pt]
&\to \int \frac{\d^3 k}{(2\pi)^3} \frac{[(\dgalO)^2](\k) [(\dgalO)^2](-\k)+4 \sigma^2(2 [(\dgalO)](\k) [(\dgalO)](-\k) -  P(k) ) }{P(k)+N^2 } \label{eq:app_f22_map}\,,
\end{align}
where we used~(\ref{eq:delta2}) to get  
\beq
-\frac{P(k)}{P(k)+N^2} [(\dgalO)^2](\k)+ [(\tilde b_1 \star \dgalO)^2](\k) \ \to\ \frac{N^2}{P(k) +N^2} [(\dgalO)^2](\k) \, .
\eeq
Repeating the same steps for $b_n$ and $\fnleq$---using $\tilde b_n(\k) \approx - b_n P(k)/(P(k)+N^2)$ and $\tilde b_1(\k) \supset n \times (n! \sigma^{2n-2}) b_n^2 $---we recover the same Fisher matrix as in the case without cosmic variance, but with $N^2 \to P(k) +N^2$: 
\beq
\begin{aligned}
F_{n,m} &\ =\  \int  \frac{\d^3 k}{(2\pi)^3}  \frac{[(\dgalO)^n](\k) [(\dgalO)^m](-\k) + n \delta_{n,m} n! \sigma^{2n-2} (2|\dgalO(\k)|^2 - \sigma^2)}{P(k) + N^2} \,,\\[4pt]
F_{n, \rm eq} &\ =\   \sum_m \int  \frac{\d^3 k}{(2\pi)^3} \, m \bar b_m \frac{[(\dgalO)^n](\k)[(\dgalO)^{m-1}\delta_{\rm NG}[\dgalO] ](-\k)}{P(k)+N^2}   \\
& \quad\ \ +  \frac{2}{3} \delta_{n,2} \int \frac{\d^3 p}{(2\pi)^3} \frac{B_{\rm eq}(\p,\k-\p\hskip 1pt)}{(P(|\k-\p\hskip 1pt |)+N^2) (P(k)+N^2)^2} (2|\dgalO(\k)|^2 - \sigma^2)\, \\[6pt]
F_{\rm eq, eq} &\ =\ \int  \frac{\d^3 k}{(2\pi)^3}  \frac{\delta_{\rm NG}[\dgalO](\k) \hskip 2pt \delta_{\rm NG}[\dgalO](-\k)}{P(k)+N^2} \\
& \quad\ \ +  \frac{1}{9} \int \frac{\d^3 p}{(2\pi)^3} \frac{B_{\rm eq}(\p,\k-\p\hskip 1pt) B_{\rm eq}(-\p, \k)}{(P(|\k-\p\hskip 1pt |)+N^2) (P(p)+N^2) (P(k)+N^2)^2} (2|\dgalO(\k)|^2 - \sigma^2)\, .
\end{aligned}
\eeq
For simplicity, we only keep the leading contributions in powers of $\dgalO$ and neglect terms that vanish for high signal-to-noise.

\vskip 4pt
The above Fisher matrix elements depend on the specific realization of the fields.  
As before,
we can use the ergodic theorem to replace this with the statistical averages:
\beq\label{eq:app_F_final}
\begin{aligned}
F_{n,m} &\ \to\ V  \int   \frac{\d^3 k}{(2\pi)^3}  \frac{  \langle [ (\dgalO)^n](\k) [(\dgalO)^m](-\k) \rangle' + n \delta_{n,m} n! \sigma^{2n}}{P(k) + N^2} \,,\\[12pt]
F_{n, \rm eq} &\ \to\  V \sum_m \int  \frac{\d^3 k}{(2\pi)^3} \, m \bar b_m \frac{\langle [(\dgalO)^n](\k)[(\dgalO)^{m-1}\delta_{\rm NG}[\dgalO] ](-\k) \rangle'}{P(k)+N^2}   \\
& \quad\ \ \ + \frac{2}{3}\delta_{n,2} \int \frac{\d^3 p}{(2\pi)^3} \frac{B_{\rm eq}(\p,\k-\p\hskip 1pt)}{(P(|\k-\p\hskip 1pt |)+N^2) (P(k)+N^2)} \ , \\[12pt]  
F_{\rm eq, eq} &\ \to\ V \int  \frac{\d^3 k}{(2\pi)^3}  \frac{\langle \delta_{\rm NG}[\dgalO](\k) \delta_{\rm NG}[\dgalO](-\k)\rangle' }{P(k)+N^2}\\
&\quad\ \ \ +  \frac{V}{9} \int \frac{\d^3 p}{(2\pi)^3} \frac{B_{\rm eq}(\p,\k-\p\hskip 1pt ) B_{\rm eq}(-\p, \k)}{(P(|\k-\p\hskip 1pt |)+N^2) (P(p)+N^2) (P(k)+N^2)} \, ,
\end{aligned}
\eeq
where $V$ is the survey volume.

\subsection{Map Versus Bispectrum}\label{app:map_bi}

Our final task is  to show that the map-level and bispectrum-only constraints on $b_2$ and $\fnleq$ are equivalent, which  turns out to be the same as the argument that the bispectrum provides an optimal estimator for $\fnleq$ (see e.g.~\cite{Babich:2005en,Creminelli:2006gc}).  

\vskip 4pt
We want to prove that the bispectrum Fisher matrix element (setting $N\to 0$ for simplicity),
\beq\label{eq:bi_fish_app}
F^{\rm (B)}_{\rm eq,eq} = \frac{V}{6} \int  \frac{\d^3 k  \hskip 2pt \d^3 p \hskip 2pt  \d^3 q }{(2\pi)^9}   \frac{B_{\rm eq}(\p,\q\hskip 1pt) B_{\rm eq}(-\p,-\q\hskip 1pt)}{P(k)  P(q)  P(p) }  (2\pi)^3 \delta_{\rm D}(\p +\q-\k)  \,,
\eeq
is the same as the map-level result $F_{\rm eq, eq}^{\rm (M)}$ in~(\ref{eq:app_F_final}).  
Inserting the definition of $\delta_{\rm NG}(\k)$ in (\ref{eq:dNG}) and (\ref{equ:PhiNG}) into~(\ref{eq:app_F_final}), we get
\begin{align}
F_{\rm eq, eq}^{\rm (M)} &= \frac{V}{18} \int  \frac{\d^3 k \hskip 2pt \d^3 p  \hskip 2pt \d^3 q}{(2\pi)^9}\,  \frac{B_{\rm eq}(\p,\q\hskip 1pt) \big[ B_{\rm eq}(-\p,-\q\hskip 1pt)+ 2  B_{\rm eq}(-\p, \k) \big]}{P(k) P(q) P(p)} (2\pi)^3 \delta_{\rm D}(\p +\q-\k)\, .
\label{equ:A25}
\end{align}
By construction, the bispectra $B_{\rm eq}(\p,\q\hskip 1pt)$ are invariant under permutations of $\k$, $\p$ and $\q$.  By a permutation of the second term in (\ref{equ:A25}), we therefore get
\begin{align}
F_{\rm eq, eq}^{\rm (M)} &= \frac{V}{6} \int  \frac{\d^3 k  \hskip 2pt \d^3 p  \hskip 2pt \d^3 q}{(2\pi)^9}  \frac{B_{\rm eq}(\p,\q \hskip 1pt) B_{\rm eq}(-\p,-\q \hskip 1pt)}{P(k) P(q) P(p)} \,  (2\pi)^3 \delta_{\rm D}(\p +\q-\k) \, ,
\end{align}
which agrees precisely with the bispectrum result (\ref{eq:bi_fish_app}).  

\vskip 4pt
A similar comparison applies for $b_2$, where the associated bispectrum is given by (\ref{equ:B2}). 
Substituting this into~(\ref{eq:bi_fish_app}), the bispectrum Fisher matrix becomes
\begin{align}
F^{\rm (B)}_{2,2} &=  \frac{V}{6} \int  \frac{\d^3 k_1 \d^3 k_2  \d^3 k_3 }{(2\pi)^9}   \frac{4 \left(P(k_1)P(k_2)+P(k_1)P(k_3)+P(k_2)P(k_3)\right)^2}{P(k_1)  P(k_2)  P(k_3) }  (2\pi)^3 \delta_{\rm D}(\k_1 +\k_2+\k_3) \nonumber \\
&= V\int  \frac{\d^3 k_1 \d^3 k_2  }{(2\pi)^9}  \left( \frac{2 P(k_1)P(k_2)}{P(k_3) }  + 4 P(k_1) \right) .
\end{align}
For comparison, evaluating the correlators in the map-level Fisher matrix~(\ref{eq:app_f22_map}), we get 
\begin{align}
F_{2,2}^{\rm (M)} &= V \int \frac{\d^3 k}{(2\pi)^3} \frac{[(\dgalO)^2](\k) [(\dgalO)^2](-\k)+4 \sigma^2(2 [(\dgalO)](\k) [(\dgalO)](-\k) -  P(k) ) }{P(k) } \nonumber \\
&= V \int \frac{\d^3 k \hskip 2pt \d^3 p}{(2\pi)^6} \left(\frac{2 P(p) P(|\k-\p\hskip 1pt |)}{P(k)} + 4 P(p)\right) , 
\end{align}
where we used $\sigma^2 = \int \d^3 p \,P(p) /(2\pi)^3$. After relabeling the integration variables,  it is easy to see that the map-level and bispectrum Fisher matrix elements 
are indeed the same.

\clearpage
\phantomsection
\addcontentsline{toc}{section}{References}
\bibliographystyle{utphys}
\bibliography{LSSrefs}

\end{document}